\documentclass[aps,prd,preprint,showpacs,showkeys,nofootinbib,superscriptaddress]{revtex4-1}
\pdfoutput=1
\usepackage{amsmath}
\usepackage{amssymb}
\usepackage{dcolumn}
\usepackage{bm}
\usepackage{multirow}
\usepackage{blkarray}
\usepackage{bbold}
\usepackage{feynmp}                                     
\usepackage{hyperref,graphicx}           
\usepackage{caption}
\usepackage{subcaption}
\usepackage{xcolor}
\usepackage[utf8]{inputenc}
\usepackage{slashed}

\newcommand{\omp}{\hat{P}}
\newcommand{\lmm}{\hat{V}}
\newcommand{\opp}{\hat{S}}

\newcommand{\constmass}{\mu_{\psi}}

\begin{document}

\title{Light Hidden Mesons through the $Z$ Portal}

\author{Hsin-Chia Cheng}
\affiliation{Center for Quantum Mathematics and Physics (QMAP), Department of Physics,\\ University of California, Davis, CA 95616, USA}
\author{Lingfeng Li}
\affiliation{Jockey Club Institute for Advanced Study, Hong Kong University of Science and Technology, Clear Water Bay, Kowloon, Hong Kong}
\author{Ennio Salvioni}
\affiliation{Physik-Department, Technische Universit\"at M\"unchen, 85748 Garching, Germany}
\author{Christopher B. Verhaaren}
\affiliation{Center for Quantum Mathematics and Physics (QMAP), Department of Physics,\\ University of California, Davis, CA 95616, USA}

\begin{abstract}
Confining hidden sectors are an attractive possibility for physics beyond the Standard Model (SM). They are especially motivated by neutral naturalness theories, which reconcile the lightness of the Higgs with the strong constraints on colored top partners. We study hidden QCD with one light quark flavor, coupled to the SM via effective operators suppressed by the mass $M$ of new electroweak-charged particles. This effective field theory is inspired by a new tripled top model of supersymmetric neutral naturalness. The hidden sector is accessed primarily via the $Z$ and Higgs portals, which also mediate the decays of the hidden mesons back to SM particles. We find that exotic $Z$ decays at the LHC and future $Z$ factories provide the strongest sensitivity to this scenario, and we outline a wide array of searches. For a larger hidden confinement scale $\Lambda\sim O(10)\;\mathrm{GeV}$, the exotic $Z$ decays dominantly produce final states with two hidden mesons. ATLAS and CMS can probe their prompt decays up to $M\sim 3\;\mathrm{TeV}$ at the high luminosity phase, while a TeraZ factory would extend the reach up to $M\sim 20\;\mathrm{TeV}$ through a combination of searches for prompt and displaced signals. For smaller $\Lambda \sim O(1)\;\mathrm{GeV}$, the $Z$ decays to the hidden sector produce jets of hidden mesons, which are long-lived. LHCb will be a powerful probe of these emerging jets. Furthermore, the light hidden vector meson could be detected by proposed dark photon searches.

\end{abstract}
\preprint{TUM-HEP-1203-19}


\maketitle
\thispagestyle{empty}
\tableofcontents
\setcounter{page}{1} 
\section{Introduction\label{sec.intro}}
A confining hidden sector, or ``hidden valley,'' that interacts weakly with the visible sector is an intriguing possibility for physics beyond the Standard Model (BSM)~\cite{Strassler:2006im}. There is, however, a wide range of options for such theories, and without some guiding principle it is unclear what particles and interactions should be expected to populate the hidden valley. One way to make progress is to connect the hidden sector to the resolution of one or more of the open questions of the SM, such as the naturalness and dark matter problems.

The naturalness problem of the weak scale remains one of the most important motivations to explore BSM physics. However, the apparent lack of new physics (NP) signals at the Large Hadron Collider (LHC) has put significant strain on ``traditional'' models of supersymmetry (SUSY) and of composite Higgs. In these models light colored top partners cancel the quadratic ultraviolet (UV) sensitivity of the Higgs mass parameter. Except for some special cases, the LHC bounds on new colored particles have reached beyond $1$~TeV. A possible way to evade the strong experimental constraints is that the new light particles interact only weakly with the SM sector. Consequently, there has been growing interest in constructing and studying theories of ``neutral naturalness'' (NN), where the top partners do not carry SM color quantum numbers. Many different models of NN have been proposed. They can be classified by the spin and gauge quantum numbers of the top partners, which can be fermions~\cite{Chacko:2005pe,Cai:2008au,Poland:2008ev,Batell:2015aha,Serra:2017poj,Csaki:2017jby,Serra:2019omd} or scalars~\cite{Burdman:2006tz,Cheng:2018gvu,Cohen:2018mgv}, and carry SM electroweak (EW) quantum numbers~\cite{Burdman:2006tz,Cai:2008au,Poland:2008ev,Serra:2019omd} or be complete SM singlets~\cite{Chacko:2005pe,Batell:2015aha,Serra:2017poj,Csaki:2017jby,Cheng:2018gvu,Cohen:2018mgv}. 

A common feature of NN models is that the top partners are charged under a hidden color gauge group, whose coupling is approximately equal to the SM strong coupling at high scales. This preserves the relation between the top-Higgs and top-partner-Higgs couplings, enabling the cancelation of the respective leading contributions to the Higgs potential. As a result, the hidden sector is expected to confine and the hidden hadrons are often important in NN phenomenology~\cite{Craig:2015pha}. In this way NN models provide welcome guidance in the vast hidden valley parameter space, by singling out representative scenarios and setting well-motivated targets for experimental searches. 

In the presence of a hidden strong gauge group two regimes are possible, depending on whether light matter fields are present, which allow hidden color strings to break. If all matter fields charged under the hidden gauge group are heavier than the confinement scale $\Lambda$, once these particles are pair produced the gauge-flux string that connects them cannot break. This ``quirky'' scenario~\cite{Kang:2008ea} occurs in many NN models, and the related signatures were explored previously~\cite{Burdman:2008ek,Harnik:2008ax,Fok:2011yc,Harnik:2011mv,Chacko:2015fbc}. On the other hand, if there are matter fields with masses below $\Lambda$, then pair production of hidden-colored particles results in final states containing light hidden hadrons. These may be produced via parton showers if the event energy is much larger than the confinement scale. In this paper, we focus on this second scenario.

As the hidden color coupling is linked to the SM strong coupling, the confinement scales of the two sectors are also related. When the hidden sector has fewer light states its color gauge coupling runs faster, resulting in a confinement scale somewhat larger than that of SM QCD. In this case $\Lambda$, which for light constituents sets the masses of the hidden hadrons, typically ranges from a few hundred MeV to a few tens of GeV. To satisfy experimental constraints, the constituents must be mostly singlets under the SM gauge interactions. However, NN models often predict additional heavy states that carry both hidden color and SM EW quantum numbers. After EW symmetry breaking the doublets and singlets can mix, resulting in small couplings between the $Z$ and Higgs ($h$) bosons and the light constituents. As a consequence, the light hidden hadrons are produced in rare $Z$ and $h$ decays. The associated phenomenology is the main subject of this paper. 

A concrete NN example that leads to the above scenario is a realization of the recently-proposed tripled top (TT) framework~\cite{Cheng:2018gvu}, which guides our discussion. This model naturally contains both the light singlet fermions that confine into hidden hadrons, and the TeV scale EW-charged fermions that mix with them. However, we emphasize that these necessary ingredients are fairly typical expectations of NN theories. For example, in the Twin Higgs framework some twin quarks can be lighter than the confinement scale, while hidden-colored, SM EW-charged fermions do appear at the (multi-)TeV scale in non-SUSY UV completions~\cite{Chacko:2005pe,Geller:2014kta,Barbieri:2015lqa,Low:2015nqa,Cheng:2015buv,Cheng:2016uqk}. Furthermore, such scenario may also arise in hidden valley theories motivated by other open problems of the SM. 

We minimize model dependence by phrasing our discussion within an effective field theory (EFT), where the low-energy effects of the heavy EW-charged particles are captured by higher-dimensional operators built out of the SM and light hidden fields. The prototypical hidden sector we consider contains one Dirac fermion $\psi$ with mass $m_\psi$, transforming in the fundamental representation of a hidden $SU(N_d)$ color group that confines at scale \mbox{$\Lambda \gg m_\psi$}, where we take $N_d = 3$ as motivated by NN. The fermion is a complete singlet under the SM gauge symmetries, but interacts with the visible sector according to the Lagrangian $\mathcal{L} = \mathcal{L}_{\rm SM} +  \overline{\psi} ( i \slashed{D} - m_\psi ) \psi  - \hat{G}_{\mu\nu}^a \hat{G}^{a\,\mu\nu}/4 + \mathcal{L}_6$, where the covariant derivative acting on $\psi$ is $D_\mu = \partial_\mu - i g_d \hat{G}_\mu^a t^a$. For simplicity we assume $CP$ conservation in the hidden sector, and therefore take $m_\psi$ to be real and neglect the $\theta$-term of hidden QCD. The non-renormalizable part of the Lagrangian reads
\begin{equation}\label{eq:EFT}
\mathcal{L}_6 = \frac{m_t^2}{M^2 v^2} \Big( |H|^2\, \overline{\psi}_R i \slashed{D} \psi_R + \mathrm{h.c.} + i (D_\mu H)^\dagger H\, \overline{\psi}_R \gamma^\mu \psi_R + \mathrm{h.c.}  +  c_g\, \frac{\alpha_d}{12\pi} |H|^2  \hat{G}_{\mu\nu}^a \hat{G}^{a\,\mu\nu} \Big),
\end{equation}
where the $m_t^2/v^2 = y_t^2 /2$ factor manifests its origin from a solution to the naturalness problem. $M$ is the mass of heavy EW-charged fermions, whereas $c_g$ is a dimensionless parameter ($4\pi \alpha_d \equiv g_d^2$). The EFT description encoded by $\mathcal{L}$ is valid at energies \mbox{$\Lambda \ll E \ll M$}. The chiral structure of $\mathcal{L}_6$ is inspired by the TT model, and will be assumed in the rest of the paper.\footnote{The effective operators in Eq.~\eqref{eq:EFT} can be contrasted, for example, with those obtained in a Fraternal Twin Higgs (FTH) model~\cite{Craig:2015pha} with light twin $b$, $\mathcal{L}_6^{\rm FTH} = |H|^2 \big( m_{\hat{b}}\,\overline{\hat{b}} \hat{b} - \alpha_d \hat{G}_{\mu\nu}^a \hat{G}^{a\,\mu\nu} / (12\pi) \big) / f^2 $ with $ m_{\hat{b}} = y_{\hat{b}} f / \sqrt{2}$.} In unitary gauge, the second operator in Eq.~\eqref{eq:EFT} yields a small coupling of $\psi$ to the $Z$ boson,
\begin{equation}  \label{eq:Zpsipsi_coupling}
\frac{g_Z}{2}\, \frac{m_t^2}{M^2}\, \overline{\psi}_{R} \gamma^\mu \psi_{R} Z_\mu ,
\end{equation}
where $g_Z \equiv \sqrt{g^2 + g^{\prime\,2}}$. The first operator in Eq.~\eqref{eq:EFT} can be rewritten, by using the leading-order equation of motion for $\psi$, as $m_t^2  |H|^2 m_\psi \overline{\psi} \psi / (M^2 v^2)$. Hence, the couplings of the Higgs to hidden particles read
\begin{equation} \label{eq:hpsipsi_hgg_coupling}
\frac{m_t^2}{M^2} \frac{h}{v} \Big( m_\psi\,  \overline{\psi} \psi  +  c_g\, \frac{\alpha_d}{12\pi} \hat{G}_{\mu\nu}^a \hat{G}^{a\,\mu\nu} \Big) \,.
\end{equation}
The interactions in Eqs.~\eqref{eq:Zpsipsi_coupling} and \eqref{eq:hpsipsi_hgg_coupling} mediate decays of the $Z$ and $h$ to the hidden sector. In addition, they control the decays of the lowest-lying hidden hadrons, which are light mesons~\cite{Farchioni:2007dw}. We focus on a $1$-flavor hidden QCD theory because this case arises most naturally in the TT model. Since the anomaly removes all chiral symmetries and therefore no light pseudo-Nambu-Goldstone bosons (pNGBs) are expected~\cite{Creutz:2006ts}, several among the lightest mesons play important roles in the phenomenology. This is in contrast to the multi-flavor scenario, where the hidden pions are expected to dominate, and whose phenomenology will be the subject of a separate publication~\cite{DarkPions}. Incidentally, we note that the lightest baryon of the $1$-flavor theory, $(\psi \psi \psi)$ with spin $3/2$, could be cosmologically stable due to hidden baryon number conservation and provide an interesting candidate for asymmetric dark matter, along the lines followed in Ref.~\cite{Garcia:2015toa} for the FTH model with light twin bottom.\footnote{See also Ref.~\cite{Farina:2015uea} for asymmetric dark matter in the mirror TH model, and Ref.~\cite{Terning:2019hgj} for the FTH with $\hat{b}$ much heavier than the confinement scale.}

An indirect constraint on the EFT in Eq.~\eqref{eq:EFT} comes from $1$-loop corrections to the $T$ parameter of electroweak precision tests (EWPT). Diagrams with two insertions of the second operator give a quadratically divergent contribution that we expect to be cut off at $M$, resulting in $\widehat{T} = \kappa N_d y_t^2 m_t^2 / (16\pi^2 M^2)$, with $\kappa$ a UV-dependent $O(1)$ coefficient. For example, the fermionic sector of the TT model gives $\kappa = 4/3$, as shown in Appendix~\ref{app:Tparameter}. The current constraint $\widehat{T}\lesssim 10^{-3}$ then bounds $M \gtrsim 0.87\;\mathrm{TeV}$, while future $e^+ e^-$ colliders will be able to improve the sensitivity to $\widehat{T} \lesssim 10^{-4}$ \cite{Fan:2014vta}, corresponding to $M \gtrsim 2.7$~TeV.\footnote{In Ref.~\cite{Fan:2014vta} a thorough study of the future reach of EWPT was performed, considering several $e^+ e^-$ collider proposals. While the bound on $\widehat{T}$ quoted here suffices as a rough estimate for our purposes, we caution that a precise assessment requires the detailed analysis presented there.} As usual, though, the EWPT constraints can be importantly affected by additional unknown corrections. In this paper we focus on {\it direct\/} probes of the hidden sector, which, as we show, extend the reach to larger $M$ in many regions of parameter space.

Identifying the most promising signatures requires a detailed understanding of the spectrum and decay patterns of the lightest hidden mesons. Depending on their masses and on the mass scale of the heavy EW-charged particles, the expected signals range from prompt two-body decays, to hidden parton showers followed by displaced decays. We analyze many of these possibilities in detail, finding that $Z$ decays, especially, will have an impressive NP reach both at the LHC and at future $e^+ e^-$ colliders. While rare and exotic Higgs decays have been extensively studied at the LHC (see e.g. Refs.~\cite{Strassler:2006ri,Curtin:2013fra}) and also at future Higgs factories~\cite{Liu:2016zki,Alipour-Fard:2018lsf}, the sensitivity of $Z$ decays to hidden sectors has been much less explored. Reference~\cite{Blinov:2017dtk} studied the LHC reach on both prompt and long-lived decays for a hidden Abelian Higgs model, whereas Ref.~\cite{Liu:2017zdh} focused on prompt decays at future $Z$ factories in scenarios where the hidden sector contains the dark matter particle.
 
This paper is organized as follows. In Sec.~\ref{sec:model} we introduce a new version of the TT framework~\cite{Cheng:2018gvu}, which realizes the scenario outlined above. While the model provides important motivation, our discussion is structured so that readers whose primary interest is phenomenology may omit Sec.~\ref{sec:model}. In Sec.~\ref{sec:light_resonances} we discuss the essential ingredients for our phenomenological study: the production of the light hidden mesons through $Z$ and Higgs decays, as well as the expected pattern of hidden meson lifetimes and branching ratios. Section~\ref{sec:pheno} presents the analysis of the collider phenomenology, and contains our main results. We summarize and conclude in Sec.~\ref{sec:conclusions}. Finally, three Appendices complete the paper.

\section{A New Tripled Top Model}\label{sec:model}
This section presents a NN model with a confining hidden sector of light mesons, whose constituents are SM-singlet fermions. The interactions between the hidden and visible sectors are described by the EFT in Eq.~\eqref{eq:EFT}. The construction is an alternative realization of the TT framework proposed in Ref.~\cite{Cheng:2018gvu}.

Tripled top models are supersymmetric extensions of the SM which include two copies of a hidden top sector, each charged under its own $SU(3)$ color gauge group. The hidden sectors consist of vector-like $SU(2)_L$-doublet and -singlet hidden top supermultiplets, and the stabilization of the Higgs mass is achieved by means of an accidental supersymmetry in their spectrum~\cite{Burdman:2006tz,Cheng:2018gvu}. In Ref.~\cite{Cheng:2018gvu} the scalar top partners were chosen to be complete SM singlets. However, from the point of view of naturalness there is no particular preference for EW-singlet top partners. It is straightforward to write down an alternative model where the roles of doublets and singlets are switched.\footnote{Retaining the same notation for the fields, this amounts to exchanging $u^c_{B,C} \leftrightarrow Q_{B,C}$ and $u^\prime_{B,C} \leftrightarrow Q^{\prime c}_{B,C}$ in Ref.~\cite{Cheng:2018gvu}.} The superpotential of the three top sectors is then
\begin{align}
W_{Z_3} \,& =\, y_t \left( Q_A H u^c_A + Q_B H u^c_B + Q_C H u^c_C\right)  +  M (Q_B Q_B^{\prime c} + Q_C Q_C^{\prime c} )  + \omega  ( u^\prime_B u_B^c  + u^\prime_C u_C^c ) \,,
\end{align}
where $H = H_u$ and the subscript $A$ labels the SM fields, while $B$ and $C$ denote the two hidden sectors. A $Z_3$ symmetry is assumed to relate the top Yukawa couplings and the $SU(3)$ gauge couplings of the three sectors. It is softly broken to a $Z_2$ that exchanges the $B$ and $C$ sectors by the supersymmetric mass terms $M$ and $\omega$. The scale $M$ is taken to be multi-TeV, while the size of $\omega$ will be discussed momentarily. We have neglected the additional superpotential terms $\sim \bar{y} (Q_B^{\prime c} H_d u^\prime_B + \{ B \to C \})$, as they constitute a hard breaking of the $Z_3$. The SM fields have the usual charges under the EW $SU(2)_L \times U(1)_Y$,
\begin{equation}
H = \begin{pmatrix} h^+ \\ h^0 \end{pmatrix} \sim \mathbf{2}_{1/2}\,, \qquad Q_{A} = \begin{pmatrix} t_{A} \\ b_{A} \end{pmatrix} \sim \mathbf{2}_{1/6}\,, \qquad u_A^c \sim \mathbf{1}_{-2/3}\,,
\end{equation}
which also defines the component fields. The charges of the $B$ and $C$ fields are chosen to be
\begin{equation} 
Q_{B,C} = \begin{pmatrix} t_{B,C} \\ b_{B,C} \end{pmatrix} \sim \mathbf{2}_{- 1/2}\,, \qquad Q^{\prime c}_{B,C} = \begin{pmatrix} b^{ \prime c}_{B,C} \\ t^{ \prime c}_{B,C} \end{pmatrix} \sim \mathbf{2}_{1/2}\,,\qquad u^c_{B,C},\, u^\prime_{B,C} \sim \mathbf{1}_0\,.
\end{equation}
In the above expression the ``$u$'' fields are $SU(2)_L$ singlets, while ``$t$'' states are the electrically-neutral components of doublets. The hypercharges are chosen such that ``$u$'' fields in the $B,C$ sectors are complete SM singlets.  In addition, the following form is assumed for the leading soft SUSY-breaking masses,
\begin{equation} \label{e.softmass}
V_{\rm s} = \widetilde{m}^2 ( | \widetilde{Q}_A |^2 + \left| \tilde{u}^{c}_A  \right|^2 ) - \widetilde{m}^2 ( | \widetilde{Q}_B |^2 + | \widetilde{Q}_{C} |^2 ) \,.
\end{equation}
The soft mass $\widetilde{m}$ is assumed to be close to $M$, so that the colored $A$ stops are raised to the multi-TeV scale. On the other hand, the cancelation between $M^2$ and $\widetilde{m}^2$ makes the hidden sector scalars $\widetilde{Q}_{B,C}$ light, with masses
\begin{equation} \label{eq:tuning}
\Delta \equiv \sqrt{M^2 - \widetilde{m}^2}\, \ll  M
\end{equation}
in the few hundred GeV range. The Higgs potential in this new model is {\it identical} to the one presented in Ref.~\cite{Cheng:2018gvu}. However, here the light EW-doublet scalars $\widetilde{Q}_{B,C}$ play the role of the top partners, cutting off the quadratic contribution to the Higgs potential from the top quark loop. For this reason, we call them ``top siblings.'' In addition, the supermultiplets $u'_{B,C}, \, u^c_{B,C}$, which are denoted as ``top cousins,'' are complete SM singlets, hence $\omega$ can be taken very small without violating any experimental constraint. The scalar components $ \tilde{u}'_{B,C} ,\, \tilde{u}^c_{B,C}$ are still expected to receive sizable soft SUSY-breaking masses and become heavy. Conversely, the fermions remain light and, if $\omega$ is smaller than the confinement scale \mbox{$\Lambda_{\mathrm{QCD}_{B, C}} \equiv \Lambda$} of $SU(3)_{B,C}$, they efficiently break the hidden QCD strings and form light hadrons. This is the region of parameters we are interested in: a TT model with light singlet cousin fermions, which for brevity we simply call TT in this work.

As described, this setup successfully stabilizes the Higgs mass against the multi-TeV scale $M$. Yet, for it to be a complete natural theory in the UV, the peculiar pattern of opposite-sign, equal-magnitude soft mass terms in Eq.~\eqref{e.softmass} must be explained, as well as the proximity of the soft-breaking and SUSY masses in Eq.~\eqref{eq:tuning}. A possible origin of the special structure of soft masses was presented in Ref.~\cite{Cheng:2018gvu}, whereas $\Delta \ll M$ requires a $\sim \Delta^2 / M^2$ fine-tuning in the absence of a theoretical mechanism that relates the soft and SUSY masses. We do not discuss these issues any further here, since our purpose is to use the model as an example for phenomenological studies. We also note that Eq.~(\ref{e.softmass}) only represents the leading soft SUSY-breaking terms in the top sector. The $A$ sector gluino and light generation squarks must also have multi-TeV SUSY-breaking masses to satisfy LHC bounds. All other fields can receive subleading SUSY-breaking masses of a few hundred GeV which split the fermions and bosons in the supermultiplets, without spoiling naturalness.  

For each of the two hidden sectors, by integrating out the heavy fields with masses $\sim M$ and $\sim \Delta$ we obtain a $1$-flavor QCD with couplings to the SM dictated by Eq.~\eqref{eq:EFT}. However, 
to explicitly demonstrate how the results arise from a UV-complete model, we keep 
the heavy states ``integrated in'' in the following discussion.
Since the two hidden sectors are identical,  we only discuss the $B$ sector. For simplicity, we assume the Higgs sector is in the decoupling limit at large $\tan\beta$, so in unitary gauge $h^0 = \langle h^0 \rangle + h / \sqrt{2}$ where $\langle h^0 \rangle = v / \sqrt{2}$, $v\simeq 246\;\mathrm{GeV}$, and $h$ denotes the physical Higgs boson. The mass matrix for the fermions is 
\begin{equation}
- \begin{pmatrix} u^\prime_B & t_B \end{pmatrix} \mathcal{M}_F \begin{pmatrix} u_B^c \\ t_B^{ \prime c} \end{pmatrix}, \qquad \mathcal{M}_F =  \begin{pmatrix} \omega & 0 \\ m_t & M \end{pmatrix},
\end{equation}
where $m_t = y_t \langle h^0 \rangle$. It is diagonalized by $R(\theta_L)^T \mathcal{M}_F\, R(\theta_R) = \mathrm{diag} \,(m_\psi, M_{\Psi^0})$, where the rotations are given by (we use capital letters for the mass eigenstate fields)
\begin{equation}
\begin{pmatrix} u^\prime_B \\ t_B \end{pmatrix} \to R(\theta_L) \begin{pmatrix} U^\prime_B \\ T_B \end{pmatrix}, \qquad \begin{pmatrix} u_B^c \\ t_B^{\prime c} \end{pmatrix} \to R(\theta_R) \begin{pmatrix} U_B^c \\ T_B^{\prime c} \end{pmatrix},  \qquad R(\theta) \equiv \begin{pmatrix} \cos\theta & \sin\theta \\ - \sin\theta & \cos\theta \end{pmatrix} 
\end{equation}
with mixing angles
\begin{equation} \label{eq:mix_angles}
 \sin \theta_L = \frac{m_\psi}{M}\, \sin\theta_R \simeq \frac{ m_t  \omega}{M^2 + m_t^2} \,, \qquad \sin \theta_R \simeq \frac{m_t}{\sqrt{M^2 + m_t^2}} \,. 
\end{equation}
The first equality in Eq.~\eqref{eq:mix_angles} is exact whereas the others have been expanded for small $\omega$. As a result, $ \psi_B \equiv (U^\prime_B, U_B^{c\,\dagger})$ form a Dirac fermion with small mass of $O(\omega)$, whereas $\Psi_B^0 \equiv (T_B, T_B^{\prime c\,\dagger})$ form a Dirac fermion with large mass of $O(M)$,
\begin{equation}\label{eq:fermion_masses}
m_\psi \simeq \frac{M  \omega}{\sqrt{M^2 + m_t^2}}\,\,, \qquad M_{\Psi^0} \simeq \sqrt{M^2 + m_t^2}\,,
\end{equation}
where we have expanded for small $\omega$. The electrically-charged $\Psi_B^- \equiv (b_B, b_B^{\prime c\,\dagger})$ form a Dirac fermion with $Q =- 1$ and mass $M$.\footnote{The hard $Z_3$-breaking superpotential $-\, \bar{y} (Q_B^{\prime c} H_d u^\prime_B + \{ B \to C \})$ would modify $\mathcal{M}_F$ to $\begin{pmatrix} \omega & \overline{m}  \\ m_t & M \end{pmatrix}$, where $\overline{m} = \bar{y} v \cos\beta/\sqrt{2}$, $m_t = y_t v \sin\beta / \sqrt{2}$ and we have kept $\tan\beta$ arbitrary. If $\omega \ll \overline{m} \ll m_t, M$ we would find $\sin\theta_L \simeq  M \overline{m} / (M^2 + m_t^2)$ and $m_\psi \simeq m_t \overline{m} / \sqrt{M^2 + m_t^2}$, whereas $\theta_R$ and $M_{\Psi^0}$ remain as in Eqs.~\eqref{eq:mix_angles} and \eqref{eq:fermion_masses}. In this work we set $\bar{y} = 0$.}

The mixing parameterized by $R(\theta_R)$ couples $\psi_B$ to the $Z$ boson. In four-component spinor notation the coupling reads $(g_Z / 2) \sin^2 \theta_R\, \overline{\psi}_{BR} \slashed{Z} \psi_{BR}\,$, which to leading order in a large $M$ expansion gives Eq.~\eqref{eq:Zpsipsi_coupling}. Similarly, the $R(\theta_L)$ mixing matrix leads to a $\overline{\psi}_{BL} \slashed{Z} \psi_{BL}$ coupling, but this is suppressed by an extra factor of $(\omega / M)^2$, so we neglect it. After rotating to the fermion mass eigenstate basis, the top Yukawa interactions couple the light eigenstate to the Higgs boson as $(y_t / \sqrt{2})\, \sin\theta_L \cos\theta_R\,h\, \overline{\psi}_B \psi_B\,$. Expanding in large $M$ gives the first term in Eq.~\eqref{eq:hpsipsi_hgg_coupling}.

We calculate the Higgs coupling to hidden gluons by recalling that, given a set of Dirac fermions $f$ and complex scalars $\phi$ which transform in the fundamental of $SU(N_d)$ and with couplings $- \mathcal{L} = \sum_f g_{hff} h \bar{f} f  + \sum_{\phi} g_{h\phi \phi} h \phi^\ast \phi$, the $1$-loop Higgs coupling to gluons reads, allowing for off-shell Higgs with four-momentum $p_h^\mu\,$,
\begin{equation} \label{eq:hgg_general}
 \frac{\alpha_{d}}{16\pi} \Bigg[\frac{4}{3} \sum_{f} \frac{g_{hff}}{m_f} A_{1/2} \Big(\frac{p_h^2}{4m_f^2}\Big) + \frac{1}{6} \sum_{\phi} \frac{g_{h\phi\phi}}{m^2_\phi} A_{0} \Big(\frac{p_h^2}{4m_\phi^2}\Big) \Bigg] \hat{G}_{\mu\nu}^{a}\hat{G}^{a\,\mu\nu} h \,.
\end{equation}
Here $A_{1/2}(\tau) = 3 [\tau + (\tau - 1) f(\tau)]/ (2\tau^2)$ and $A_0 (\tau) = 3 [f(\tau) - \tau] / \tau^2\,,$ with
\begin{equation}
f(\tau) = \left\{
     \begin{array}{lr}
       \arcsin^{2}\sqrt{\tau}\,, & \quad \tau\leq 1\,,\\
       -\frac{1}{4}\left[\log\left(\frac{1+\sqrt{1-1/\tau}}{1-\sqrt{1-1/\tau}}\right)-i\pi\right]^{2}\,, & \quad\,\, \tau > 1\,.
     \end{array}
   \right.
\end{equation}
In the fermionic term the relevant couplings are $g_{h\psi_B \psi_B} = - y_t \sin\theta_L \cos\theta_R  / \sqrt{2}$ and $g_{h \Psi^0_B \Psi^0_B} = y_t \sin\theta_R \cos\theta_L / \sqrt{2}\,$. Since $A_{1/2} (\tau) \sim - 3 \log^2 \tau / (8\tau)\,[1]$ at $\tau \to \infty\, [\tau \to 0]$, for small $\omega$ the contribution of $\psi_B$ can be neglected, yielding $(c_g)_{\mathrm{fermions}} \simeq 1$ in Eq.~\eqref{eq:hpsipsi_hgg_coupling}. 

Finally, to calculate the second term in Eq.~\eqref{eq:hgg_general} we must discuss the scalar sector. The mass matrices are
\begin{equation} \label{eq:Bscalarmasses_newmodel}  
- \begin{pmatrix} \tilde{u}^\prime_B & \tilde{t}_B \end{pmatrix}^\ast  \mathcal{M}^2_{S}  \begin{pmatrix} \tilde{u}^\prime_B \\ \tilde{t}_B \end{pmatrix},   \qquad - \begin{pmatrix}  \tilde{u}_B^c & \tilde{t}^{\,\prime c}_B \end{pmatrix}  \mathcal{M}^2_{S^c}  \begin{pmatrix} \tilde{u}_B^c \\ \tilde{t}_B^{\,\prime c}  \end{pmatrix}^\ast ,
\end{equation}
with
\begin{equation} \label{eq:Bscalarmasses_newmodel_2}
\mathcal{M}^2_S = \begin{pmatrix} \omega^2 & m_t \omega \\ m_t \omega & \Delta^2 + m_t^2 \end{pmatrix} + \delta m^2 \mathbb{1}_2, \qquad \mathcal{M}^2_{S^c} =  \begin{pmatrix} \omega^2 + m_t^2 & m_t M \\ m_t M & M^2 \end{pmatrix} + \delta m^2 \mathbb{1}_2 ,
\end{equation}
where the terms proportional to $\delta m^2$ include in a crude way the effects of subleading SUSY-breaking masses. We expect $\delta m^2 \sim (100\;\mathrm{GeV})^2 \ll \Delta^2, M^2$. Diagonalization is achieved through $R (\phi_L)^T \mathcal{M}^2_{S} R(\phi_L) = \mathrm{diag} \,(m_{U^\prime}^2,  M_T^2)$ and $R(\phi_R)^T \mathcal{M}^2_{S^c} R(\phi_R) = \mathrm{diag}\,(m_{U}^2 , M_{T^\prime}^2)$, with rotations
\begin{equation}
\begin{pmatrix} \tilde{u}^\prime_B \\ \tilde{t}_B \end{pmatrix} \to R(\phi_L) \begin{pmatrix} \widetilde{U}^\prime_B \\ \widetilde{T}_B \end{pmatrix},\qquad \begin{pmatrix} \tilde{u}_B^c \\ \tilde{t}_B^{\,\prime c} \end{pmatrix} \to R(\phi_R) \begin{pmatrix} \widetilde{U}_B^c \\ \widetilde{T}_B^{\prime c} \end{pmatrix} 
\end{equation}
and mixing angles
\begin{equation}
\sin\phi_L \simeq \frac{ m_t \omega}{\Delta^2 + m_t^2} \, ,\qquad \phi_R = \theta_R \,.
\end{equation}
The physical masses are
\begin{equation} \label{eq:scalar_masses}
m^2_{U^\prime} \simeq \frac{\Delta^2 \omega^2}{\Delta^2 + m_t^2} + \delta m^2\, \,,\quad M^2_T \simeq \Delta^2 + m_t^2\,, \quad m^2_{U} =  m_\psi^2 + \delta m^2 \,,\quad M^2_{T^\prime} \simeq M_{\Psi^0}^2\,.
\end{equation}
In addition we have $\tilde{b}_B$ and $\tilde{b}_B^{\prime c}$ with charges $-1$ and $+1$, respectively, and mass $\Delta$. From the $D$-term potential we obtain the couplings to the Higgs,
\begin{align}
g_{h \widetilde{U}^\prime_B \widetilde{U}^\prime_B} &\,=\, \sqrt{2} \,y_t \sin\phi_L (m_t \sin\phi_L - \omega \cos \phi_L), \,\,\, \;\;\; g_{h \widetilde{T}_B \widetilde{T}_B} =  \sqrt{2} \,y_t \cos \phi_L (m_t \cos \phi_L + \omega \sin \phi_L), \nonumber \\
g_{h \widetilde{U}_B^c  \widetilde{U}_B^c} &\,=\, \sqrt{2} \,y_t \cos\phi_R (m_t \cos\phi_R - M \sin \phi_R), \;\, g_{h \widetilde{T}_B^{\prime c} \widetilde{T}_B^{\prime c}} =  \sqrt{2} \,y_t \sin \phi_R (m_t \sin \phi_R + M \cos \phi_R) .
\end{align}
As in the fermion case, at small $\omega$ the contribution of the light scalars to Eq.~\eqref{eq:hgg_general} can be neglected. The leading term originates from the $\widetilde{T}_B$, yielding $(c_g)_{\rm scalars} \simeq M^2 / (4\Delta^2)$ in Eq.~\eqref{eq:hpsipsi_hgg_coupling}. Since $\Delta \ll M$ requires a $\sim \Delta^2/M^2$ accidental cancellation, a moderate tuning of $O(10)\%$ corresponds to $(c_g)_{\rm scalars} \sim 2\,$-$\,4$. In passing, we note that the current constraint on the $T$ parameter only requires $\Delta \gtrsim 400\;\mathrm{GeV}$, as shown in Appendix~\ref{app:Tparameter}, making a study of the collider phenomenology of the EW-doublet scalar top partners an interesting direction for future work. In this paper, however, we concentrate on the less model-dependent production of hidden hadrons from decays of SM particles.

\section{Production and Decays of the Light Hidden Mesons}\label{sec:light_resonances}
This section sets the stage for our study of the hidden sector phenomenology. We first discuss production of the light hidden mesons through rare $Z$ and Higgs decays, and then analyze the expected pattern of hidden meson lifetimes and branching ratios. 

\subsection{Production}
The coupling in Eq.~\eqref{eq:Zpsipsi_coupling} gives the $Z$ a width for decay to one hidden sector,
\begin{equation} \label{ZFermions}
\Gamma(Z \to \overline{\psi} \psi) \simeq \frac{N_d \,g_Z^2}{96\pi}  \frac{m_t^4}{M^4} \,m_Z \left(1 - \frac{m_\psi^2}{m_Z^2} \right) \left(1 - \frac{4m_\psi^2}{m_Z^2} \right)^{1/2}.
\end{equation}
Taking the hidden color factor $N_d = 3\,$, the corresponding branching ratio into both sectors is
\begin{equation}\label{eq:ZBr}
\text{BR}(Z \to \overline{\psi}_{B,C} \psi_{B,C}) \approx 2.2\times 10^{-5} \left( \frac{ 2\,\mathrm{TeV}}{M}\right)^4 .
\end{equation}
From Eq.~\eqref{eq:hpsipsi_hgg_coupling} we calculate the widths for Higgs decay to the light hidden fermions and gluons,
\begin{equation} \label{hFermions}
\Gamma(h \to \overline{\psi} \psi) \simeq \frac{N_d y_t^2}{16\pi}  \frac{m_\psi^2 m_t^2 }{M^4}  \,m_h \left(1 - \frac{4m_\psi^2}{m_h^2} \right)^{3/2} , \qquad \Gamma(h \to \hat{g} \hat{g}) = \frac{\alpha_d^2 m_h^3}{72\pi^3 v^2} \frac{m_t^4}{M^4}\,c_g^2 .
\end{equation}
The corresponding branching ratios
\begin{align}
\text{BR}(h \to \overline{\psi}_{B,C} \psi_{B,C}) &\,\approx 1.6\times 10^{-6} \left( \frac{m_\psi}{0.5 \, \mathrm{GeV}}\right)^2  \left( \frac{ 2\,\mathrm{TeV}}{M}\right)^4 \,, \nonumber \\
\text{BR}(h \to \hat{g}_{B,C} \hat{g}_{B,C}) &\,\approx 2.0\times 10^{-4}  \left( \frac{\alpha_d}{0.18}\right)^2  \left( \frac{ 2\,\mathrm{TeV}}{M}\right)^4  \left( \frac{ c_g }{4}\right)^2 ,
\label{eq:hBr}
\end{align}
show that Higgs decays to hidden fermions are negligible. Including $2$-loop running with one flavor, the coupling is $\alpha_d = \alpha_d (m_h/2; \Lambda) \simeq 0.18$ for $\Lambda = 5\;\mathrm{GeV}$,\footnote{We have $ \alpha_d (m_h/2; \Lambda) \simeq 0.12\,(0.24)$ for $\Lambda = 1\,(10)\;\mathrm{GeV}$.} and we have chosen as reference $c_g = 4$, motivated by the TT discussion in Sec.~\ref{sec:model}. Equations~\eqref{eq:ZBr} and \eqref{eq:hBr} assume the existence of two identical hidden sectors (labeled $B$ and $C$), as in the TT model; unless otherwise noted, we retain this assumption throughout the paper. The results for a scenario with a single hidden sector are trivially obtained by adjusting appropriate factors of $2$.

At the LHC, the inclusive cross section for $Z$ production is
\begin{equation} \label{eq:Zinclusive}
\sigma (pp \to Z) = K_Z  \frac{\pi^2 \alpha_Z}{N_c s} \sum_q (v_q^2 + a_q^2)\, L_{q\bar{q}} \left(\frac{m_Z^2}{s} \right) \stackrel{\sqrt{s} \,=\, 13\,(14)\;\mathrm{TeV}}{\approx} 54.5\,(58.9)\;\mathrm{nb}\,,
\end{equation}
where $a_{f} = T_{Lf}^3$ and $v_{f}=a_{f} - 2 s_w^2 Q_{f}$, while $K_Z = 1.3$ is an approximate $K$-factor that accounts for QCD corrections~\cite{Aad:2016naf}. The parton luminosity is $L_{q\bar{q}} (\tau) = \int_{\tau}^{1} \frac{dx}{x} \big[ f_q (x) f_{\bar{q}}(\tau/x) + f_q (\tau/x) f_{\bar{q}}(x) \big]$ and in numerical evaluations we use MSTW2008NLO PDFs~\cite{Martin:2009iq} with factorization scale set to $m_Z/2$. The Higgs cross section is a thousand times smaller, $\sigma_h \approx 48.6\,(54.7)\;\mathrm{pb}$ at $13\,(14)\;\mathrm{TeV}$ for the dominant gluon fusion channel~\cite{Cepeda:2019klc}. Combining these with Eqs.~\eqref{eq:ZBr} and \eqref{eq:hBr} we find that the expected number of $Z$ decays to the hidden sectors is $\sim 120\, (0.18 / \alpha_d)^2 (4/c_g)^2$ times larger than the analogous number for the Higgs. Turning to future electron-positron colliders, the total cross section for production at the $Z$ pole is
\begin{equation} \label{eq:Zpole_xsec}
\sigma(e^+ e^-\to Z) = K_{\rm QED}  \frac{12 \pi}{m_Z^2} \,\mathrm{BR}(Z\to ee) \approx 43.6\, \mathrm{nb}\,,
\end{equation}
where $K_{\rm QED} \simeq 0.73$ accounts for QED photon radiation~\cite{ALEPH:2005ab}. A $Z$-factory will be able to produce $10^{9}$ to $10^{12}$ $Z$ bosons, corresponding to an integrated luminosity of  $22.9\; \mathrm{fb}^{-1} ( \mathrm{ab}^{-1})$ for the GigaZ~(TeraZ) option. On the other hand, a Higgs factory running at $\sqrt{s} \sim 240\,$-$250$~GeV will yield a lower number of Higgses, ranging from $10^6$ to $10^7$ depending on the collider configuration. In light of these considerations, in what follows we focus on $Z$ decays. Nonetheless, we discuss Higgs decays when they provide a useful term of comparison.

\subsection{Decays}
For $\omega < \Lambda$, the color group in each hidden sector has one light quark flavor. This theory does not predict a light pNGB: using standard $J^{PC}$ notation, the lightest hadrons are expected to be the $s$-wave $0^{-+}, 1^{--}$ mesons, and the $p$-wave $0^{++}$ meson, which we denote as $\omp, \lmm,$  and $\opp$, respectively. Lattice calculations have not yet provided precise information about the mass spectrum~\cite{Farchioni:2007dw}, in particular no attempt to evaluate $m_{\lmm}$ has been made. We make the reasonable assumption that $m_{\omp} \lesssim m_{\lmm} < m_{\opp}$, and take as reference values $m_{\omp}, m_{\lmm} = 2\Lambda$,\footnote{This choice is motivated by the spectrum of SM QCD, where $\Lambda_{\mathrm{QCD}_A} \approx 370$~MeV according to a $2$-loop RG analysis~\cite{Cheng:2018gvu} and the mass of the $\omega$ meson is $m_\omega \approx 780\;\mathrm{MeV} \sim 2\,\Lambda_{\mathrm{QCD}_A}$.} and $\Delta m \equiv m_{\opp} - m_{\lmm} = \Lambda$. The latter is motivated by the preliminary lattice result $m_{\opp}/m_{\omp} \approx 1.5$~\cite{Farchioni:2007dw}. However, we provide results in general form, and depart from the above benchmarks whenever this has important consequences. As it will be discussed momentarily, this is especially relevant for $\Delta m$, on which the lifetime of $\opp$ depends very sensitively. The mesons decay back to SM particles through the annihilation of their constituents, which proceeds via the small couplings to the $Z$ and $h$ bosons in Eqs.~\eqref{eq:Zpsipsi_coupling} and \eqref{eq:hpsipsi_hgg_coupling}. The resulting pattern of lifetimes and branching ratios is a crucial input to study the collider phenomenology, so we analyze it in detail. By contrast, we neglect the baryons as they are significantly heavier than the lowest-lying mesons: lattice calculations place the mass of the lightest baryon at $(2.5\,$-$\,3) m_{\omp}$~\cite{Farchioni:2007dw}. 

The $\lmm \, (1^{--})$ meson decays democratically to SM fermions, through the coupling of $\psi$ to the transverse $Z$ boson in Eq.~\eqref{eq:Zpsipsi_coupling}. The width for $\lmm \to f\bar{f}$ decay is
\begin{equation} \label{eq:UpsilonWidth}
\Gamma(\lmm \to f\bar{f}) = N_d N_c^f \frac{\pi \alpha_Z^2}{12}\frac{m_t^4}{M^4} \frac{m_{\lmm}^2 |\psi(0)|^2}{m_Z^4} \frac{\left(1 - \tfrac{4m_f^2}{m_{\lmm}^2}\right)^{1/2}}{\left(1 - \tfrac{m_{\lmm}^2}{m_Z^2} \right)^2} \left[ v_f^2  \Big( 1 + \tfrac{2m_f^2}{m_{\lmm}^2} \Big) + a_f^2 \Big( 1 - \tfrac{4m_f^2}{m_{\lmm}^2} \Big) \right]\! ,
\end{equation}
where $N_c^f =3~(1)$ for quarks (leptons), and $\psi(0)$ is the wavefunction at the origin. For $\omega \ll \Lambda$ we take $|\psi(0)|^2 = \Lambda^3/(4\pi)$,\footnote{Recall that $\psi(0) = R(0)/\sqrt{4\pi}$ where $R$ is the radial wavefunction. Our simple estimate $|R(0)| = \Lambda^{3/2}$ has been checked by comparing with the decays of light SM vector mesons to $e^+ e^-$. Neglecting $Z$ exchange, defining $\Gamma_V \equiv N_c \frac{16 \pi \alpha^2}{3} \frac{|\psi(0)|^2}{m^2_V} Q_e^2$ and taking $|\psi(0)|^2 = \Lambda^3_{\mathrm{QCD}_A} / (4\pi)$ gives $\Gamma(\rho(770) \to ee) \simeq (Q_{\rm eff}^\rho)^2 \Gamma_\rho \approx 10 \;\mathrm{keV} \left( \frac{\Lambda_{\mathrm{QCD}_A}}{370\;\mathrm{MeV}} \right)^3$, $\Gamma(\omega(782) \to ee) \simeq (Q_{\rm eff}^\omega)^2 \Gamma_\omega \approx 1.1 \;\mathrm{keV} \left( \frac{\Lambda_{\mathrm{QCD}_A}}{370\;\mathrm{MeV}} \right)^3$ and $\Gamma(\phi(1020) \to ee) \simeq Q_s^2 \Gamma_\phi \approx  1.3 \;\mathrm{keV} \left( \frac{\Lambda_{\mathrm{QCD}_A}}{370\;\mathrm{MeV}} \right)^3$, where the effective charges are $Q_{\rm eff}^{\rho, \omega} = (Q_u \mp Q_d)/\sqrt{2}$. These results are in fair agreement with the measured values $\Gamma(\rho, \omega, \phi \to ee)_{\rm exp} =  \{ 7.0, 0.62, 1.3\} \;\mathrm{keV}$~\cite{Tanabashi:2018oca}.} obtaining a decay length
\begin{equation}\label{eq:tau_V}
c \tau_{\lmm} \sim 0.02   \;\mathrm{mm}\, \left( \frac{10\;\mathrm{GeV}}{m_{\lmm}} \right)^2 \left( \frac{5\;\mathrm{GeV}}{\Lambda} \right)^3 \left( \frac{M}{2\;\mathrm{TeV}} \right)^4 .
\end{equation}
We have summed over all SM fermions except the top quark, and taken as reference \mbox{$m_{\lmm} = 2 \Lambda$}. For smaller masses the decays can be displaced: taking $m_{\lmm} = 2$~GeV we find
\begin{equation}\label{eq:tau_V2}
c \tau_{\lmm} \sim 10  \;\mathrm{cm}\, \left( \frac{2\;\mathrm{GeV}}{m_{\lmm}} \right)^2 \left( \frac{1\;\mathrm{GeV}}{\Lambda} \right)^3 \left( \frac{M}{2\;\mathrm{TeV}} \right)^4 .
\end{equation}

The $\omp \,(0^{-+})$ decays dominantly to the heaviest SM fermion that is kinematically available, through exchange of the longitudinal mode of the $Z$. The corresponding width is
\begin{equation} \label{etadecay}
\Gamma(\omp \to f \bar{f}) =  N_d N_c (f)\, 2 \pi \alpha_Z^2 \frac{m_t^4}{M^4} \,a_f^2\, \frac{\constmass^2 m_f^2}{m_Z^4} \frac{|\psi (0)|^2}{m_{\omp}^2} \bigg(1 - \frac{4m_f^2}{m_{\omp}^2} \bigg)^{1/2},
\end{equation}
where we have replaced $m_\psi$ with the constituent mass $\constmass =O(1) \times \Lambda$, since the chiral symmetry breaking from the condensate dominates for $\omega \ll \Lambda$. Estimating again\footnote{The $\omp \to f \bar{f}$ decay width can also be calculated by defining the $\omp$ decay constant via $\langle 0 | \overline{\psi} \gamma^\mu \gamma_5 \psi | \omp (q) \rangle = i f_{\omp} q^\mu$. This leads to the identification $f_{\omp} = 4 \sqrt{N_d}\, \constmass | \psi (0) | / m_{\omp}^{3/2}$, which for our typical assumptions yields $f_{\omp} = \sqrt{N_d / (2\pi)}\, \constmass \approx 0.7\, \constmass$.} $|\psi(0)|^2 = \Lambda^3/(4\pi)$, for $m_{\omp} > 2 \,m_b$ we obtain a decay length
\begin{equation}\label{eq:tau_P}
c \tau_{\omp} \sim 0.3  \,\mathrm{mm} \,\left( \frac{m_{\omp}}{10\;\mathrm{GeV}} \right)^2 \left( \frac{5\;\mathrm{GeV}}{\Lambda} \right)^5 \left( \frac{\Lambda}{\constmass} \right)^2 \left( \frac{M}{2\;\mathrm{TeV}} \right)^4 ,
\end{equation}
where we took $m_{\omp} = 2\Lambda$ as reference and included subleading decays to $c$ and $\tau$. If $\omp$ is too light to decay to $b\bar{b}$, it can be long-lived. For example, taking $m_{\omp} = 2$~GeV we have
\begin{equation}
c \tau_{\omp} \sim 110  \,\mathrm{m}\, \left( \frac{m_{\omp}}{2\;\mathrm{GeV}} \right)^2 \left( \frac{1\;\mathrm{GeV}}{\Lambda} \right)^5  \left( \frac{\Lambda}{\constmass} \right)^2 \left( \frac{M}{2\;\mathrm{TeV}} \right)^4 .
\end{equation}
In this estimate we have included decays to $\mu^+ \mu^-$ as well as to $s\bar{s}$. For the latter, since $\omp \to KK$ is forbidden by $CP$ invariance the leading decay is the three-body $\omp \to KK\pi$. To approximately account for this~\cite{Dolan:2014ska,Haisch:2018kqx} we multiply the perturbative width $\Gamma(\omp \to s\bar{s})$ in Eq.~\eqref{etadecay} by $(m_{s \ast}^2 / m_s^2)(16\pi / m_{\omp}^2) \, \rho (m_K, m_K, m_\pi, m_{\omp}\,) / (1 - 4m_s^2 / m_{\omp}^2)^{1/2}$ where $\rho$ denotes the phase space for isotropic $3$-body decays~\cite{Haisch:2018kqx}, and take $m_{s\ast} = 450$~MeV as motivated by a perturbative spectator model~\cite{McKeen:2008gd}. The resulting width for decay to strange hadrons is $\sim 0.15$ of that to muons.

\begin{figure}
\includegraphics[width=0.60\textwidth]{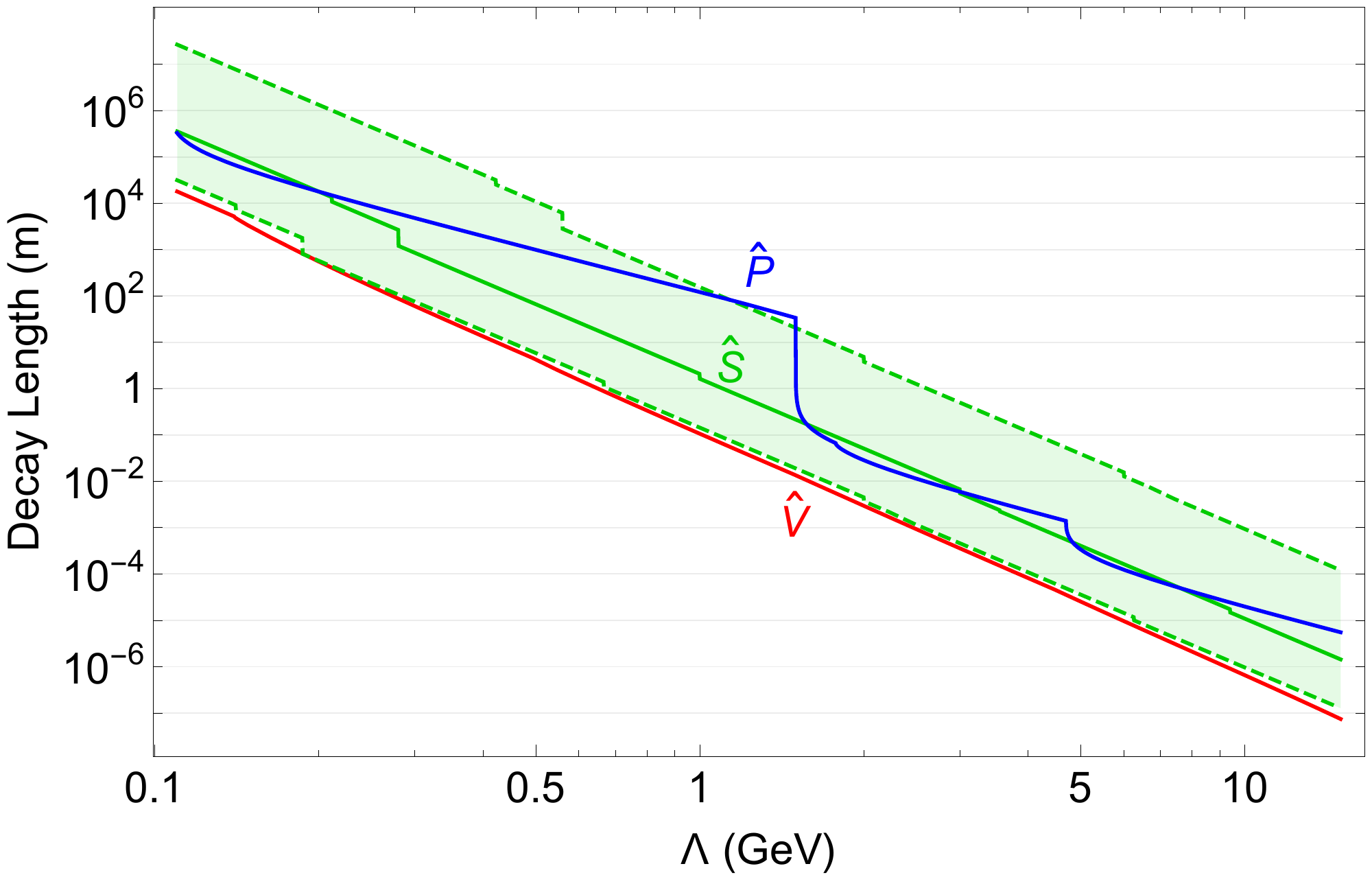} 
\caption{\label{f.mesonDecay} Decay length of the lightest hidden mesons as a function of the confinement scale. We assume $m_{\omp, \lmm} =2\Lambda$, $\mu_\psi = \Lambda$ and vary $\Lambda/2 < \Delta m < 3 \Lambda/ 2$ (green-shaded region) where $\Delta m \equiv m_{\opp} - m_{\lmm}$. The solid green curve corresponds to $\Delta m = \Lambda$. The scale $M$ is fixed to $2$~TeV; the lifetimes for different $M$ are obtained from the scaling $c\tau \propto M^4$. The plot depends very weakly on the value of $c_g\,$.}
\end{figure}

For the $p$-wave scalar $\opp \, (0^{++})$, two competing decay channels exist. The first is $\opp \to f \bar{f}$ through Higgs exchange, whose width is~\cite{Fok:2011yc}
\begin{equation}
\Gamma(\opp \to f \bar{f}) = \frac{18 N_d N^f_c}{\pi } (\lambda_{h\psi\psi} \lambda_{h f f})^2 \frac{| \psi^\prime (0) |^2}{m_h^4} \frac{\left( 1 - \frac{4m_f^2}{m_{\opp}^2} \right)^{3/2}} {\left( 1 - \frac{m_{\opp}^2}{m_h^2} \right)^{2}} \,,
\end{equation}
where $\lambda_{hff} = y_f / \sqrt{2}$. The effective $h\overline{\psi}\psi$ coupling can be estimated by observing that for $\omega \ll \Lambda$ it originates dominantly from the interaction of the Higgs to the hidden gluons in Eq.~\eqref{eq:hpsipsi_hgg_coupling}. We relate the corresponding matrix element to the $\psi$ constituent mass via the QCD trace anomaly~\cite{Shifman:1978zn}, obtaining $\lambda_{h\psi \psi} = 2c_g \constmass m_t^2 / (3b v M^2)$ where $b = 11 - 2 N_l /3$ is the leading-order coefficient of the beta function. In our case we have $N_l = 1$ light flavor. The second channel is the electric dipole-type transition \mbox{$\opp \to \lmm (Z^\ast \to f\bar{f})$}. We estimate its width by considering $\opp \to \lmm \gamma$ (which actually vanishes in our setup, since $Q_\psi =0$) and making an appropriate replacement of couplings. We begin with~\cite{Brambilla:2004wf}
\begin{equation}
\Gamma (\opp \to \lmm \gamma) = 4\, \alpha Q_\psi^2 k^3 | \varepsilon_{if} |^2 , \qquad k = \frac{m_{\opp}^2 - m^2_{\lmm}}{2m_{\opp}} = \Delta m \left( 1 - \frac{\Delta m}{2m_{\opp}} \right)\,, 
\end{equation}
where $\varepsilon_{if}$ accounts for the overlap of the radial wavefunctions of the initial and final mesons. One finds $\varepsilon_{if} \sim a$ where $a$ is the size of the bound states, hence for $\omega \ll \Lambda$ we estimate $\varepsilon_{if} \sim \Lambda^{-1}$.
The replacement of photon radiation with $Z^\ast \to f \bar{f}$ radiation is approximately captured by the substitution
\begin{equation}
\alpha Q_\psi^2 \to \left( \frac{\alpha_Z}{4} \frac{m_t^2}{M^2} \frac{k^2}{m_Z^2} \right)^2 \frac{N_f}{4\pi} \,,
\end{equation}
where $N_f$ counts the SM fermions with $2m_f < \Delta m$. Thus we obtain
\begin{equation} \label{eq:Slifetime}
\Gamma(\opp \to \lmm f \bar{f}\, ) \sim \frac{\alpha_Z^2  N_f}{16\pi} \frac{m_t^4}{M^4} \frac{k^7}{m_Z^4} | \varepsilon_{if} |^2\, .
\end{equation}
For $\omega \ll \Lambda$ the dipole decay dominates: assuming $\opp \to b\bar{b}$ is kinematically open and taking $| \psi^\prime (0) |^2 = \Lambda^5/(4\pi)$, we find
\begin{equation} \label{eq:chi_ratio}
\frac{\Gamma(\opp \to \bar{b}b)}{\Gamma(\opp \to \lmm f \bar{f}\, )} \sim \frac{c_g^2}{b^2} \frac{8 N_d N_c}{\pi N_f} \frac{y_t^2 y_b^2}{\alpha_Z^2} \frac{m_Z^4}{m_h^4}\frac{\constmass^2}{m_t^2}\frac{\Lambda^7}{k^7} \, \approx \, 10^{-5} \left( \frac{\Lambda}{5\;\mathrm{GeV}} \right)^2 \left( \frac{\constmass}{\Lambda} \right)^2 \left( \frac{\Lambda}{k} \right)^7 \left( \frac{c_g}{4} \right)^2,
\end{equation}
where $N_f = 18$ includes all SM fermions except the top and bottom. For lighter $\opp$ the ratio in Eq.~\eqref{eq:chi_ratio} is even smaller, since we need to replace $y_b$ with the Yukawas of the light fermions. Thus we expect that decays to $f\bar{f}$ can be neglected, unless $\Delta m \ll \Lambda$. When the dipole decay dominates, the $\opp$ decay length is 
\begin{equation}
c \tau_{\opp} \sim 0.1 \,\mathrm{mm}\, \left( \frac{5\;\mathrm{GeV}}{\Lambda} \right)^5 \left( \frac{\Lambda} {k}\right)^{7} \left( \frac{M}{2\;\mathrm{TeV}} \right)^4 .
\label{eq:chidecay}
\end{equation}
Fig.~\ref{f.mesonDecay} shows the $\Lambda$-dependence of the decay lengths of the light mesons for $M=2$~TeV. For a different $M$, they all scale as $M^4$.

\section{Collider Phenomenology of the Light Hidden Mesons}\label{sec:pheno}
The main production mechanism for the hidden mesons at colliders are the decays of the $Z$ and Higgs bosons. Production via decays of heavy particles in the hidden sector is also present in general, at least from the EW-charged states with mass $\sim M$ that give rise to the effective interactions in Eq.~\eqref{eq:EFT}. However, other channels may also exist, for example, in the TT model discussed in Sec.~\ref{sec:model} the scalar top partners with mass $\sim \Delta$ can also play an important role. As these additional channels are more model-dependent, being very sensitive to the detailed spectrum and decay modes, we focus on the $Z$ and $h$ decays. These probe the most interesting range of hidden meson masses, going from $\sim m_{Z,h}/2$ down to the SM QCD scale.

Very few parameters determine the phenomenology. The heavy mass $M$ controls the strength of the interactions with the SM,\footnote{As discussed in Sec.~\ref{sec:model}, in the TT model $M$ also roughly sets the mass scale of SM-colored degrees of freedom, because $\widetilde{m} \sim M$. In this case direct LHC searches for the colored stops bound $M \gtrsim 1$-$1.5\;\mathrm{TeV}$. However, this does not need to be the case in general, and a priori $M$ could be significantly smaller if only EW-charged states appear at this scale.} while the hidden confinement scale $\Lambda$ sets, modulo $O(1)$ coefficients that must calculated on the lattice, all hadronic quantities. These include the masses of the mesons (recall that we assume $\omega \ll \Lambda$, so our typical benchmark is $m_{\omp} \sim m_{\lmm} \sim 2 \Lambda$) as well as the wavefunction overlaps and the constituent mass of $\psi$. Another relevant parameter is the mass splitting $\Delta m$, on which the lifetime of the scalar meson depends very sensitively (as illustrated by Fig.~\ref{f.mesonDecay}), although it is in principle also determined by $\Lambda$. In this parameter space we have identified several different regions, each leading to distinct phenomenological predictions.

For large confinement scale, $\Lambda \gtrsim 10$~GeV, the hidden mesons are not much lighter than $m_{Z,h}/2$, hence phase space forces the $Z$ and $h$ decays to be dominantly two-body. At these larger $\Lambda$ values the mesons are likely to decay promptly back to the SM for $M\sim O(\mathrm{TeV})$. This phenomenology is discussed in Subsec.~\ref{sec:prompt}. 

If $\Lambda \gtrsim 10$~GeV but $M$ is very large, roughly $M \gtrsim 5\;\mathrm{TeV}$, then some of the mesons become long-lived. The relatively large meson masses imply that ATLAS and CMS (as well as future $Z$ factories) can be sensitive to their displaced decays, opening up an alternative strategy to detect the hidden sector. We discuss searches for heavy long-lived hidden mesons in Subsec.~\ref{sec:displaced}.

For smaller confinement scale, $\Lambda \lesssim 2\,$-$3$~GeV, the $Z$ and $h$ decays to the hidden sector result in the production of two (or more) hadronic jets, dominantly composed of light hidden mesons. In this region of parameters some or all of the hidden mesons are expected to be long-lived, so the phenomenology bears similarities with the emerging jet scenario~\cite{Schwaller:2015gea}. However, the current ATLAS and CMS emerging jet searches do not apply to our signals, due to the rather soft nature of the latter. Instead, we find that the best sensitivity is obtained at LHCb. This is discussed in Subsec.~\ref{sec:emerging}.\enlargethispage{-1cm}

If the hidden mesons are sufficiently light, then other production mechanisms become relevant, such as, for example, decays of the SM $B$ and $K$ mesons, as well as brehmsstrahlung and Drell-Yan processes. In this case the vector meson $\lmm$, which couples to the SM by mixing with the $Z$ boson, gives the largest signals. In Subsec.~\ref{sec:Bmeson} we reinterpret current bounds and future searches for dark photons in our parameter space.

Finally, for very small $\Lambda$ the hidden mesons are very long lived, and typically escape the detectors. The signals to look for are invisible $Z$ and $h$ decays. We translate the corresponding bounds to our setup in Subsec.~\ref{sec:invisible}.

\subsection{Two-body prompt decays}\label{sec:prompt}
For large values of the confinement scale, $\Lambda \sim 10$~GeV, we expect all mesons to decay promptly if $M\sim O(\mathrm{TeV})$, as shown in Fig.~\ref{f.mesonDecay}. Furthermore, the hidden mesons are sufficiently heavy that the $Z$ boson dominantly decays to two-body final states. We expect $\omp \opp$ to be the leading mode, because using an EFT for the mesons we can write the unsuppressed coupling $\hat{g}_Z Z_\mu (\opp\, \partial^\mu \omp - \omp\, \partial^\mu \opp )$.\footnote{Matching the $Z \to \omp \opp$ decay width to the one for $Z\to \overline{\psi} \psi$ in Eq.~\eqref{ZFermions}, we can identify parametrically $\hat{g}_Z \sim \sqrt{N_d }\, g_{Z\psi \psi}\,$, where $g_{Z\psi \psi} \sim g_Z \,m_t^2 / M^2$ is the $Z\overline{\psi} \psi$ coupling in Eq.~\eqref{eq:Zpsipsi_coupling}.} In our analysis we focus primarily on $Z\to (\omp \to b\bar{b}) (\opp \to \lmm f\bar{f})$ followed by $\lmm \to \ell \ell$ ($\ell = e, \mu$), which leads to $b\bar{b}\ell\ell+X$. We study this final state both at the LHC and at future $Z$-factories. At the $Z$-factories we also consider $\lmm \to \nu \bar{\nu}$, leading to $b\bar{b}\,+\mathrm{missing}\;\mathrm{momentum}+X$. This last analysis is presented in Appendix~\ref{App:bbnunu_Zfact}.

Subleading modes include decays to a (pseudo-)scalar and vector, which require one flip of the $\psi$ chirality. From the effective couplings $c_{\lmm \opp (\omp)} \hat{g}_Z \constmass\, Z_\mu \lmm^\mu \opp\,(\omp)$, with $c_{\lmm \opp (\omp)}$ dimensionless coefficients, we find $\Gamma(Z \to \lmm \opp\, (\omp)) / \Gamma(Z \to \omp \opp)  \sim \constmass^2 / (4 m_{\lmm}^2) \sim (\constmass / \Lambda)^2 c_{\lmm \opp (\omp)}^2 /16$, where the decay to longitudinally-polarized $\lmm$ is assumed to dominate. The $\lmm\omp $ mode produces final states similar to those of the dominant $\omp \opp$ mode (albeit with slightly different kinematics), so we do not discuss it further. We do analyze $\lmm \opp$, focusing on $Z\to (\lmm \to \ell\ell) (\opp \to \lmm f\bar{f})$ followed by $\lmm \to \ell^\prime \ell^\prime$. This yields a very clean $4\,$-lepton$+X$ final state, which plays an important role at the LHC. 

The last remaining two-body decay is $\lmm \lmm$, mediated by an interaction of the form $c_{\lmm \lmm} \hat{g}_Z (m_{\lmm}^2 / m_Z^2) \,\epsilon^{\mu\nu\rho\sigma} Z_\mu \lmm_\nu \partial_\rho \lmm_\sigma$ where $c_{\lmm \lmm}$ is a dimensionless coefficient. This coupling can, in general, arise from anomalous Wess-Zumino terms (see Ref.~\cite{Dror:2017nsg} for a recent discussion of light anomalous vectors). We have extracted a factor of $m_{\lmm}^2 / m_Z^2$, which is expected in our setup, to make manifest the smooth decoupling for $m_{\lmm}\ll m_Z$ of the decay width $\Gamma(Z \to \lmm \lmm) = c_{\lmm \lmm}^2 \hat{g}_Z^2 (m_{\lmm}^2 / m_Z) (1 - 4m_{\lmm}^2/m_Z^2)^{5/2} / (96\pi)$. We then find $\Gamma(Z \to \lmm \lmm) / \Gamma(Z \to \omp \opp)  \sim (m_{\lmm}/ m_Z)^2 c_{\lmm \lmm}^2 / 2\,$. For this channel we focus on the $4\,$-lepton final state $Z\to (\lmm \to \ell\ell) (\lmm \to \ell^\prime \ell^\prime)$.

We define $f_{XY}\leq 1$ as the fraction of $Z$ decays to the hidden sectors that yield the $XY$ final state. Rather than attempt to accurately estimate $f_{XY}$, based on the discussion above we simply take as reasonable benchmarks $f_{\omp \opp} \sim 1$ for the leading mode and $f_{\lmm \opp}, f_{\lmm \lmm} \sim 0.1$ for subleading modes.

\subsubsection*{$Z\to b\bar{b}\mu \mu+X$ at the LHC}
\noindent While the LHC collaborations have not yet performed a dedicated search for this final state, we can glean some information from the searches for $h \to aa\to b\bar{b}\mu\mu$, where $a$ is a light pseudoscalar~\cite{Sirunyan:2018mot,Aaboud:2018esj}. The CMS analysis~\cite{Sirunyan:2018mot} imposes softer cuts on the muons and $b$-jets, namely 
\begin{equation} \label{eq:CMS_bbmumu_cuts}
p_{T}^{\mu 1,2} > 20, 9\;\mathrm{GeV}, \qquad p_{T}^{b 1,2} > 20, 15\;\mathrm{GeV}, \qquad |\eta_{\mu, b}| < 2.4\,,
\end{equation}
and is therefore better suited to retain sensitivity to our signal than the ATLAS analysis~\cite{Aaboud:2018esj}.

To estimate the total acceptance times efficiency $(\mathcal{A}\,\epsilon)_{\rm tot}$ for our signal to pass the basic selection of Eq.~\eqref{eq:CMS_bbmumu_cuts}, we implement our model in FeynRules~\cite{Alloul:2013bka} and simulate the process $pp\to Z\to ( \omp \to b\bar{b} )(\opp  \to \lmm f\bar{f} \to \mu \mu f\bar{f})$. The $\opp \to \lmm f \bar{f}$ decay is described by an effective coupling structure $\opp\, \lmm_\mu \overline{f} \gamma^\mu P_L f$ and we take $f = u$ to capture the dominant decay to quarks. The signal is generated using MadGraph5 v2.6.6~\cite{Alwall:2014hca}, at leading order in QCD including up to one additional parton. Events are showered using Pythia8~\cite{Sjostrand:2014zea} and detector response is modeled with Delphes3~\cite{deFavereau:2013fsa}. For the latter we use the CMS card, but lower the $p_T$ threshold to $5\;\mathrm{GeV}$ for the muons and $10\;\mathrm{GeV}$ for the jets; in addition, we apply a flat total $b$-tagging efficiency $\epsilon_{bb} = 3 \times (0.4)^2 \approx 0.5$, which was estimated from the requirements described in Ref.~\cite{Sirunyan:2018mot}. With these settings we reproduce within $20\%$ the total acceptance times efficiency for the $h \to a a \to b\bar{b}\mu\mu$ signal to pass the basic selection, which is $(\mathcal{A}\,\epsilon)_{\rm tot}\sim 5\%$ for $m_{a} = 20, 40$ GeV. 

\begin{figure}[t]
\includegraphics[width=0.48\textwidth]{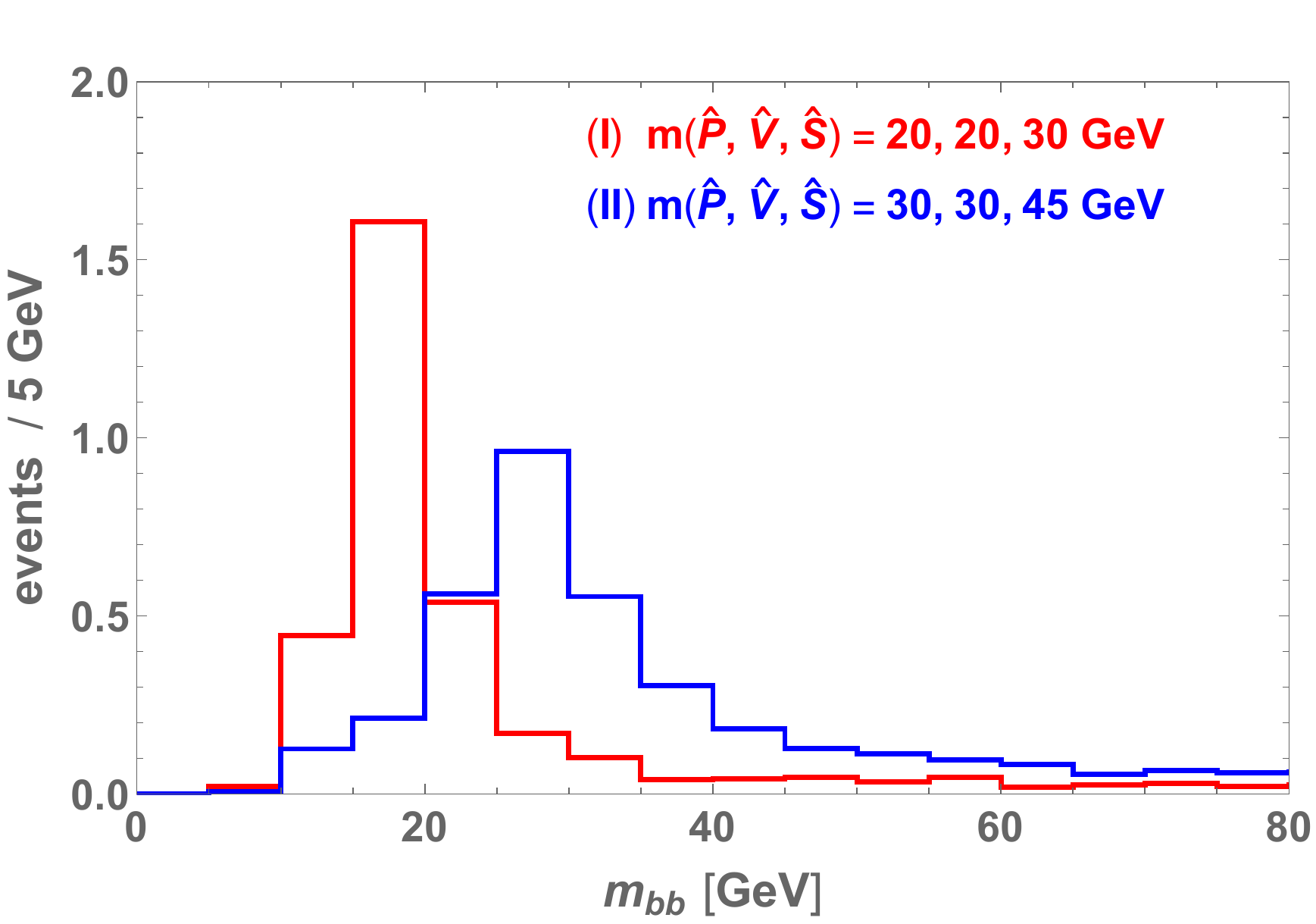}\hspace{2mm}
\includegraphics[width=0.48\textwidth]{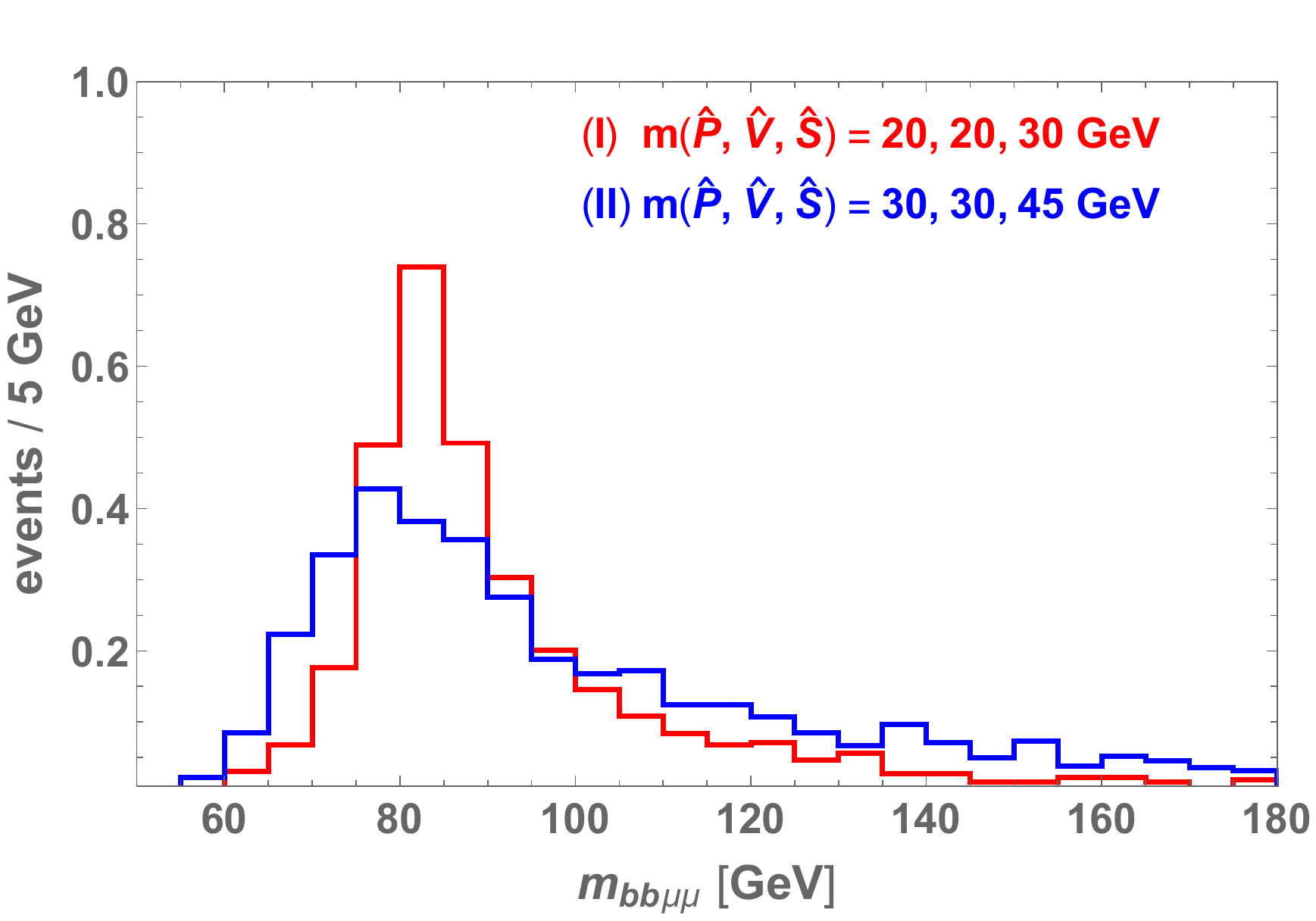}
\caption{\label{f.bbmumu_LHC}Distributions of the $b\bar{b}$ invariant mass ({\it left}) and total invariant mass of the $b\bar{b}\mu\mu$ system ({\it right}) for the $Z\to \omp \opp \to b\bar{b} \mu \mu + X$ signal at the $13$ TeV LHC, after the basic selection of Eq.~\eqref{eq:CMS_bbmumu_cuts}. The normalization corresponds to $f_{\omp \opp} = 1$ and $M = 2\;\mathrm{TeV}$. The integrated luminosity is set to $L = 35.9$ fb$^{-1}$ as in Ref.~\cite{Sirunyan:2018mot}.} 
\end{figure}

We consider two benchmark mass spectra, 
\begin{equation}
(\mathrm{I}) \quad m_{\omp,\, \lmm,\, \opp} = 20, 20, 30\;\mathrm{GeV}, \qquad\quad (\mathrm{II}) \quad m_{\omp,\, \lmm, \,\opp} = 30, 30, 45\;\mathrm{GeV},
\end{equation}
representative of $\Lambda = 10\;\mathrm{GeV}$ and $15\;\mathrm{GeV}$, respectively, with $\Delta m = \Lambda$. We find $(\mathcal{A}\,\epsilon)_{\rm tot}^{\rm I} = 0.26\%$ and $(\mathcal{A}\,\epsilon)_{\rm tot}^{\rm II} = 0.30\%$, showing that the efficiency for our signal is suppressed by an extra order of magnitude compared to $h\to aa$. The expected number of signal events for a given integrated luminosity $L$ is
\begin{equation} \label{eq:signal_rate_bbmumu}
N_S = \sigma (pp\to Z)\, \mathrm{BR}(Z \to \overline{\psi}_{B,C} \psi_{B,C} ) f_{\omp \opp}\, \mathrm{BR}(\omp \to b\bar{b}) \mathrm{BR} (\lmm \to \mu\mu) (\mathcal{A}\, \epsilon)_{\rm tot} L \,,
\end{equation}
where $\sigma(pp\to Z)$ was given in Eq.~\eqref{eq:Zinclusive} and the branching ratios are $\mathrm{BR}(\omp \to b\bar{b}) \approx 0.88$ and $\mathrm{BR} (\lmm \to \mu\mu) \approx 0.034$ in this meson mass range. Note that we have taken $\mathrm{BR}(\opp \to \lmm f \bar{f}) \simeq 1$, as expected from Eq.~\eqref{eq:chi_ratio}. Assuming the integrated luminosity used in the CMS analysis, $L = 35.9$ fb$^{-1}$, we find $N_S^{\rm I,\, II} = 3.4, 3.9\, f_{\omp \opp}\, (2\;\mathrm{TeV}/M)^4$. Key distributions for the signal are shown in Fig.~\ref{f.bbmumu_LHC}. 

To estimate the reach we exploit the fact that the CMS paper provides (see their Fig.~3, middle-right panel) the expected background yields after the basic selection, down to $m_{bb\mu\mu} = 75\;\mathrm{GeV}$. We find that the best sensitivity is achieved by selecting $m_{bb\mu\mu} \in [75, 90]\,\mathrm{GeV}$, which retains a fraction $\mathcal{A}_{bb\mu\mu}^{\rm I,\, II} = 0.50, 0.27$ of the signal. Neglecting systematic uncertainties and in the Gaussian approximation, we obtain the $95\%$ CL bounds
\begin{equation} \label{eq:Z_bbmumu_bounds}
(\mathrm{I}) \quad M \gtrsim 1.1,\, 1.4,\, 2.0\;\mathrm{TeV} \left( \frac{ f_{\omp \opp} }{1}\right)^{1/4} , \qquad\quad (\mathrm{II}) \quad M \gtrsim 1.0,\, 1.3,\, 1.7\;\mathrm{TeV} \left( \frac{ f_{\omp \opp} }{1}\right)^{1/4} \! ,
\end{equation}
for $L = 35.9, 300$ fb$^{-1}$ at $13$~TeV and $3$ ab$^{-1}$ at $14$~TeV, respectively.

We stress that these bounds are obtained without exploiting the characteristic feature $m_{bb} \approx m_{\mu\mu} \approx m_{\omp, \lmm}$ of our signal, which would permit a further suppression of the background. 
They are, therefore, an extremely conservative illustration of the reach. We do not attempt a dedicated analysis here, but encourage the experimental collaborations to undertake it. Keeping the transverse momentum cuts as low as possible will play an important role: we have checked that softening slightly the CMS cuts to $p_{T}^{\mu 1,2} > 17, 8\;\mathrm{GeV}$ (corresponding to the thresholds for the dimuon trigger~\cite{Sirunyan:2018mot}) and $p_{T}^{b 1,2} > 15, 15\;\mathrm{GeV}$ increases the signal efficiency by a factor $1.9~(1.8)$ for benchmark I~(II). Conversely, the efficiency for our signal to pass the moderately harder cuts employed in the ATLAS selection~\cite{Aaboud:2018esj} is $5$-$10$ times smaller than for the CMS selection. 

\subsubsection*{$Z\to 4\mu+X$ at the LHC}
\noindent For the $Z\to (\lmm \to \ell\ell) (\opp \to \lmm f\bar{f}\to  \ell^\prime \ell^\prime f\bar{f})$ signal we use the results of the CMS search for a light $Z^\prime$ in $Z \to 4\mu$ events~\cite{Sirunyan:2018nnz}. The basic event selection requires 
\begin{align}
4\;\mu\;\mathrm{with}\;p_T^\mu > &\,5\;\mathrm{GeV}, |\eta_\mu|<2.4,\; \mathrm{of}\;\mathrm{which} \geq 2 \;\mathrm{with}\; p_T^\mu > 10\;\mathrm{GeV}\;\mathrm{and}\,\geq 1 \;\mathrm{with}\; p_T^\mu > 20\;\mathrm{GeV}, \nonumber \\
 &\mathrm{zero}\; \mathrm{total}\;\mathrm{charge}, \qquad m_{\mu^+ \mu^-} \in [4, 120]\;\mathrm{GeV}\;\mathrm{for} \;\mathrm{all} \;\mathrm{combinations}. \label{eq:4mu_basic_sel}
\end{align}
In addition, the total invariant mass must lie within $m_{4\mu} \in [80,100]\;\mathrm{GeV}$. The $\mu^+ \mu^-$ pair with invariant mass closest to $m_Z$ is defined as $Z_1^\prime$, and the other pair as $Z^\prime_2$. Depending on its mass, the $Z^\prime$ is then typically reconstructed as $Z^\prime_1$ or $Z^\prime_2$. These requirements are not well suited to our scenario, for two reasons. First, the $m_{4\mu} > 80\;\mathrm{GeV}$ cut removes the bulk of our signal. The $f\bar{f}$ pair carries a significant amount of energy, pushing $m_{4\mu}$ well below $m_Z$. Second, the presence of {\it two} dimuon resonances causes the $Z_{1,2}^\prime$ reconstruction to either completely fail to produce a peak, as in benchmark I, or be inefficient, producing peaks at $m(Z_{1,2}^\prime) = m_{\lmm}$ but also important tails that reduce the sensitivity, as in benchmark II. Therefore, we retain the basic selection of Eq.~\eqref{eq:4mu_basic_sel}, but propose dedicated cuts to target our signal.
\begin{figure}
\includegraphics[width=0.48\textwidth]{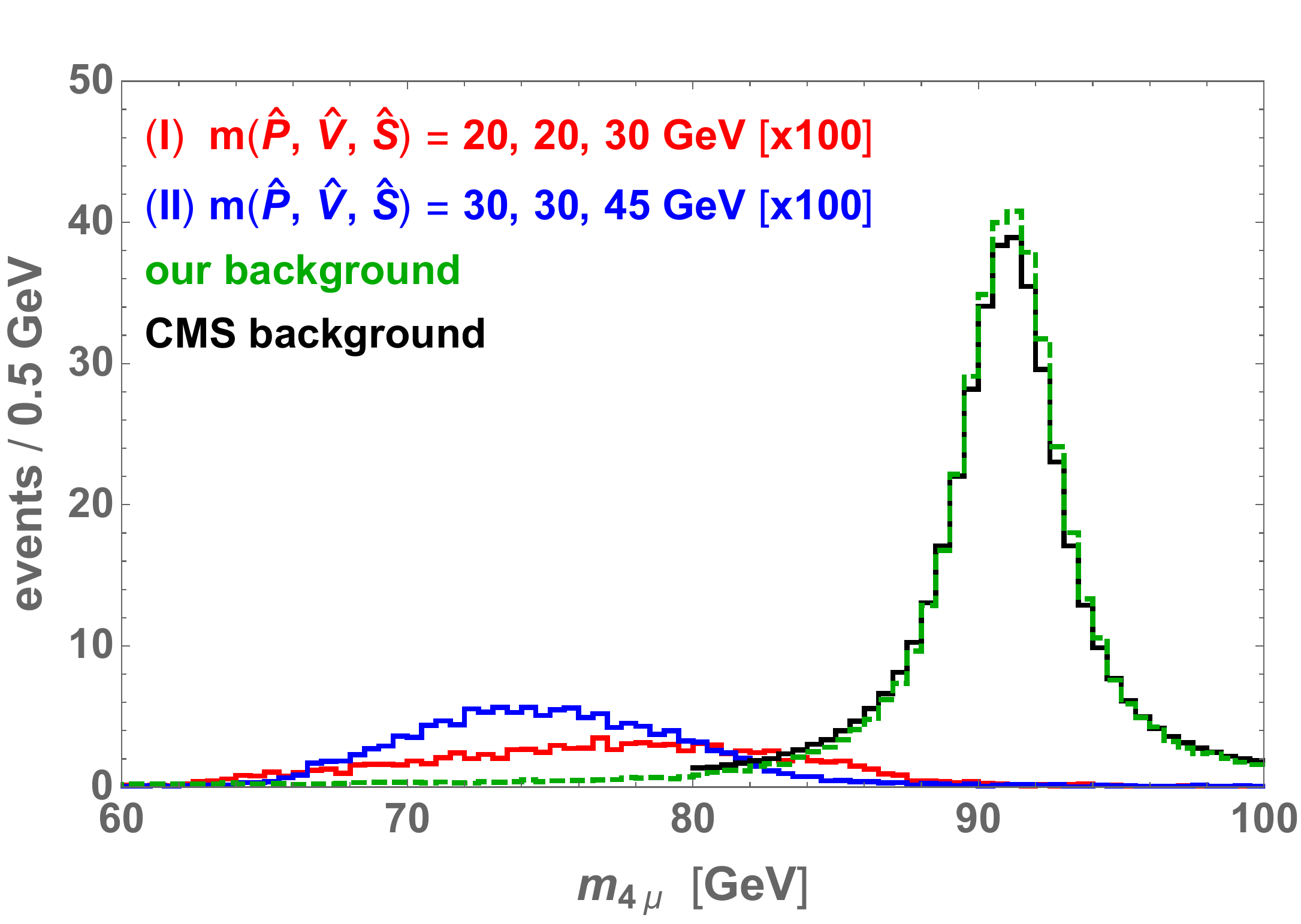}\hspace{2mm}
\includegraphics[width=0.48\textwidth]{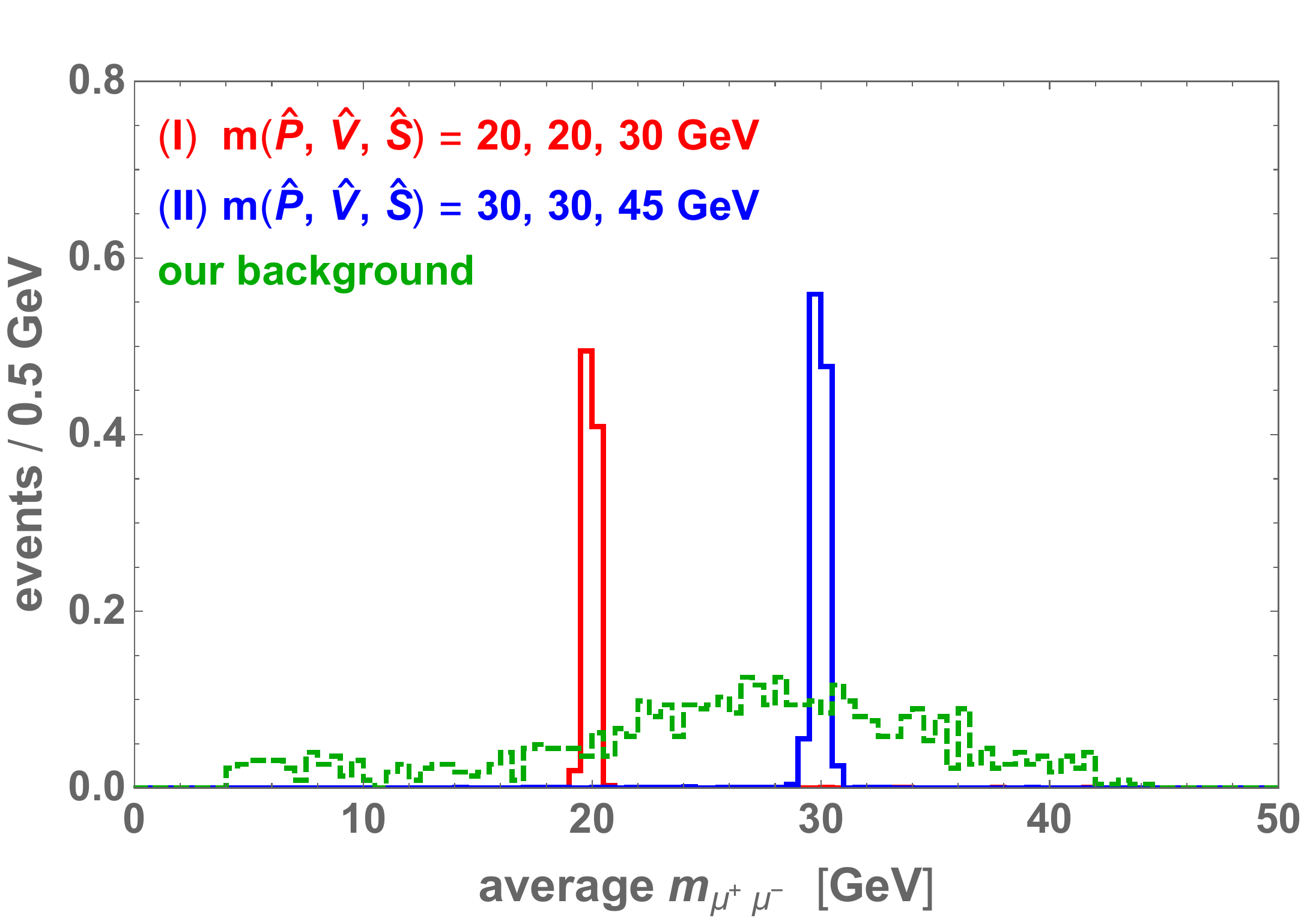}
\caption{\label{f.4mu_LHC}Distributions for the $Z\to \lmm \opp \to 4\mu + X$ signal and SM background at the $13$~TeV LHC. {\it Left:} $m_{4\mu}$ distribution after the basic selection of Eq.~\eqref{eq:4mu_basic_sel}. See the text for further details on the background prediction. {\it Right:} distribution of average $m_{\mu^+ \mu^-}$ after the additional requirements in Eq.~\eqref{eq:4mu_extra_sel}; at this stage, the background amounts to $4.0$ events. The signal normalization corresponds to $f_{\lmm \opp} = 0.1$ and $M = 2\;\mathrm{TeV}$; in the left panel it is multiplied by $100$ for the sake of illustration. We set $L = 77.3$ fb$^{-1}$ as in Ref.~\cite{Sirunyan:2018nnz}.} 
\end{figure}

The $m_{4\mu}$ distribution for the signal and SM background after the basic selection are shown in the left panel of Fig.~\ref{f.4mu_LHC}. The simulation parameters are identical to the $b\bar{b}\ell\ell + X$ analysis. For the background we generate $q\bar{q} \to 4\mu$ in the SM with up to one additional parton, including in the normalization an approximate $K$-factor of $1.3$ and a further rescaling factor of $1.15$ to match the expected number of events in the $m_{4\mu}\in [80, 100]\;\mathrm{GeV}$ window quoted by CMS~\cite{Sirunyan:2018nnz}. This simplified prescription allows us to obtain agreement at the level of $\sim 20\%$ with the shape of the background quoted by CMS, which we consider sufficient for our scope. We then require, in addition to Eq.~\eqref{eq:4mu_basic_sel}, that
\begin{equation} \label{eq:4mu_extra_sel}
m_{4\mu} \in [60, 90]\;\mathrm{GeV}, \qquad | m(\mu_1^+ \mu_1^-) - m(\mu_2^+ \mu_2^-) |\; \;\mathrm{or}\;\;  | m(\mu_1^+ \mu_2^-) - m(\mu_2^+ \mu_1^-) | < 1 \;\mathrm{GeV}.
\end{equation}
The distribution of $m_{\mu^+ \mu^-}$, averaged between the two values that are within $1\;\mathrm{GeV}$ of each other for each event that passes this additional selection, is shown for signal and background in the right panel of Fig.~\ref{f.4mu_LHC}. Finally, we require that the average $m_{\mu^+ \mu^-} \in [m_{\lmm} - 0.5\;\mathrm{GeV}, m_{\lmm} + 0.5\;\mathrm{GeV}]$, which leaves us with $N_S^{\rm I,\, II} = 0.9, 1.0\, (f_{\lmm \opp} / 0.1) (2 \;\mathrm{TeV} / M)^4$ signal and $N_B^{\rm I, \,II} = 0.1, 0.2$ background events, where we have assumed $L = 77.3$ fb$^{-1}$. Using Poisson statistics and neglecting systematics, we set the $95\%$ CL bounds
\begin{equation} \label{eq:Z_4muX_results}
(\mathrm{I}) \quad M \gtrsim 1.5,\, 2.0,\, 3.3 \;\mathrm{TeV} \left( \frac{  f_{\lmm \opp} }{0.1} \right)^{1/4} , \quad\quad (\mathrm{II}) \quad M \gtrsim 1.5,\, 2.1,\, 3.2 \;\mathrm{TeV} \left( \frac{  f_{\lmm \opp} }{0.1} \right)^{1/4} ,
\end{equation}
for $L = 77.3, 300$ fb$^{-1}$ at $13$~TeV and $3$ ab$^{-1}$ at $14$~TeV, respectively. Notice that for $M$ as large as $3.3\;\mathrm{TeV}$ the $\lmm$ and $\opp$ mesons decay promptly in this $\Lambda$ range, see Fig.~\ref{f.mesonDecay}. To conclude, we note that the $Z\to \lmm \opp$ signal discussed here shares some features with $Z\to A^\prime h_D$ studied in Ref.~\cite{Blinov:2017dtk}, where $A^\prime$ and $h_D$ are a dark photon and dark Higgs, respectively. However, in Ref.~\cite{Blinov:2017dtk} $h_D \to A^\prime A^{\prime\, (\ast)} \to 4\ell$ was selected, while remaining inclusive in the decays of the $A^\prime$. Here we have followed a different strategy, in particular we did not attemp to reconstruct the $\opp$ invariant mass peak.

The $Z\to (\lmm \to \ell\ell) (\lmm \to \ell^\prime  \ell^\prime)$ decay gives a very similar signature, except the total invariant mass of the four leptons peaks at $m_Z$. We apply the same event selection described in Eqs.~\eqref{eq:4mu_basic_sel} and \eqref{eq:4mu_extra_sel}, but modify the $m_{4\mu}$ window to $[80, 100]~\mathrm{GeV}$. For $L = 77.3$ fb$^{-1}$ this gives $10.4$ total background events, and finally selecting a narrow window around $m_{\lmm}$ we arrive at $N_S^{\rm I,\, II} = 1.3, 1.4\, (f_{\lmm \lmm} / 0.1) (2 \;\mathrm{TeV} / M)^4$ signal events and $N_B^{\rm I, \,II} = 0.2, 0.5$ background events. The resulting limits are
\begin{equation}  \label{eq:Z_4mu_results}
(\mathrm{I}) \quad M \gtrsim 1.6,\, 2.2,\, 3.4 \;\mathrm{TeV} \left( \frac{  f_{\lmm \lmm} }{0.1} \right)^{1/4} , \quad\quad (\mathrm{II}) \quad M \gtrsim 1.6,\, 2.1,\, 3.2 \;\mathrm{TeV} \left( \frac{  f_{\lmm \lmm} }{0.1} \right)^{1/4} ,
\end{equation}
for $L = 77.3, 300$ fb$^{-1}$ at $13$~TeV and $3$ ab$^{-1}$ at $14$~TeV, respectively.

\subsubsection*{$Z\to b\bar{b}\ell \ell+X$ at $Z$ factories}
\begin{figure}
\includegraphics[width=0.48\textwidth]{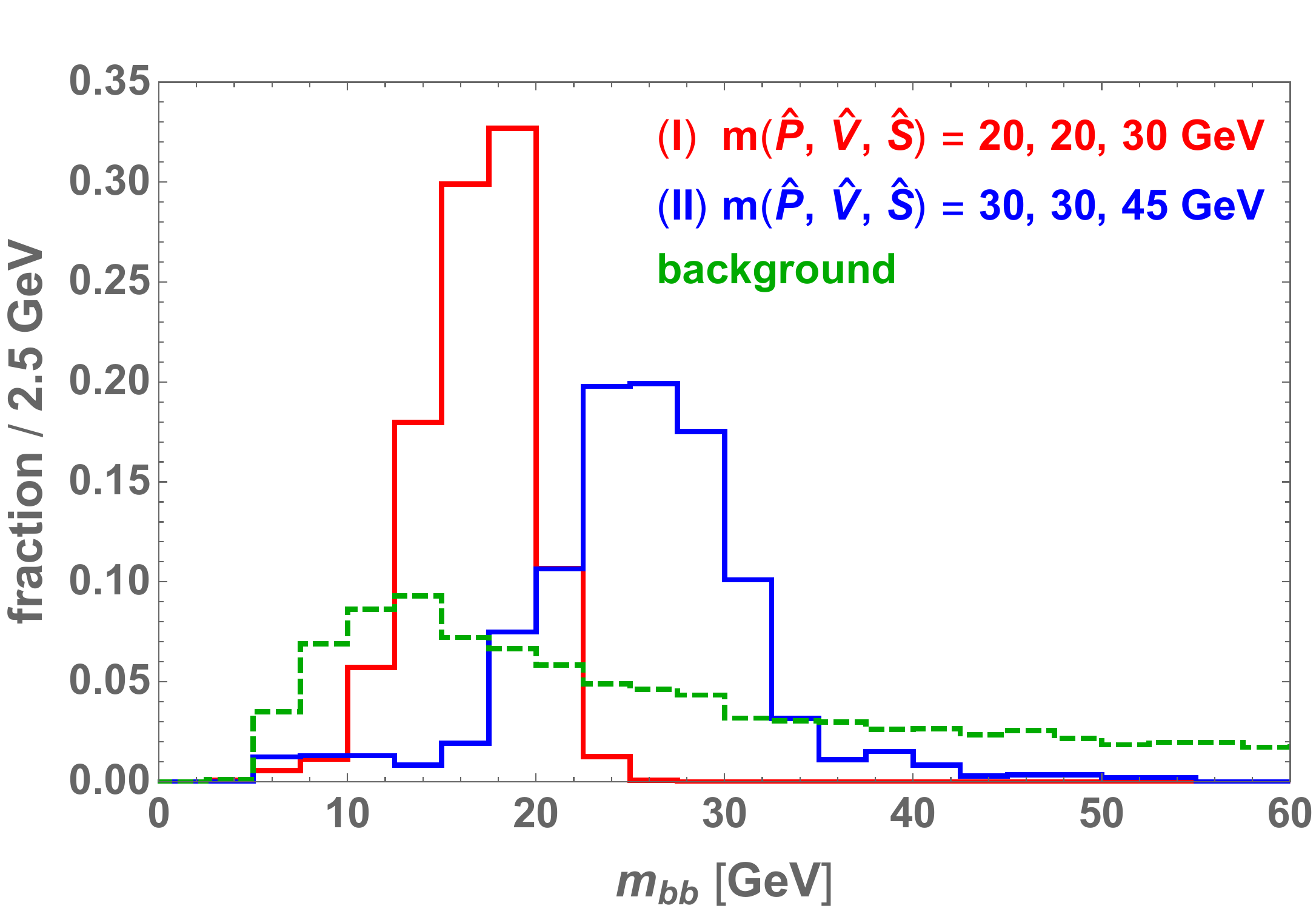}\hspace{2mm}
\includegraphics[width=0.48\textwidth]{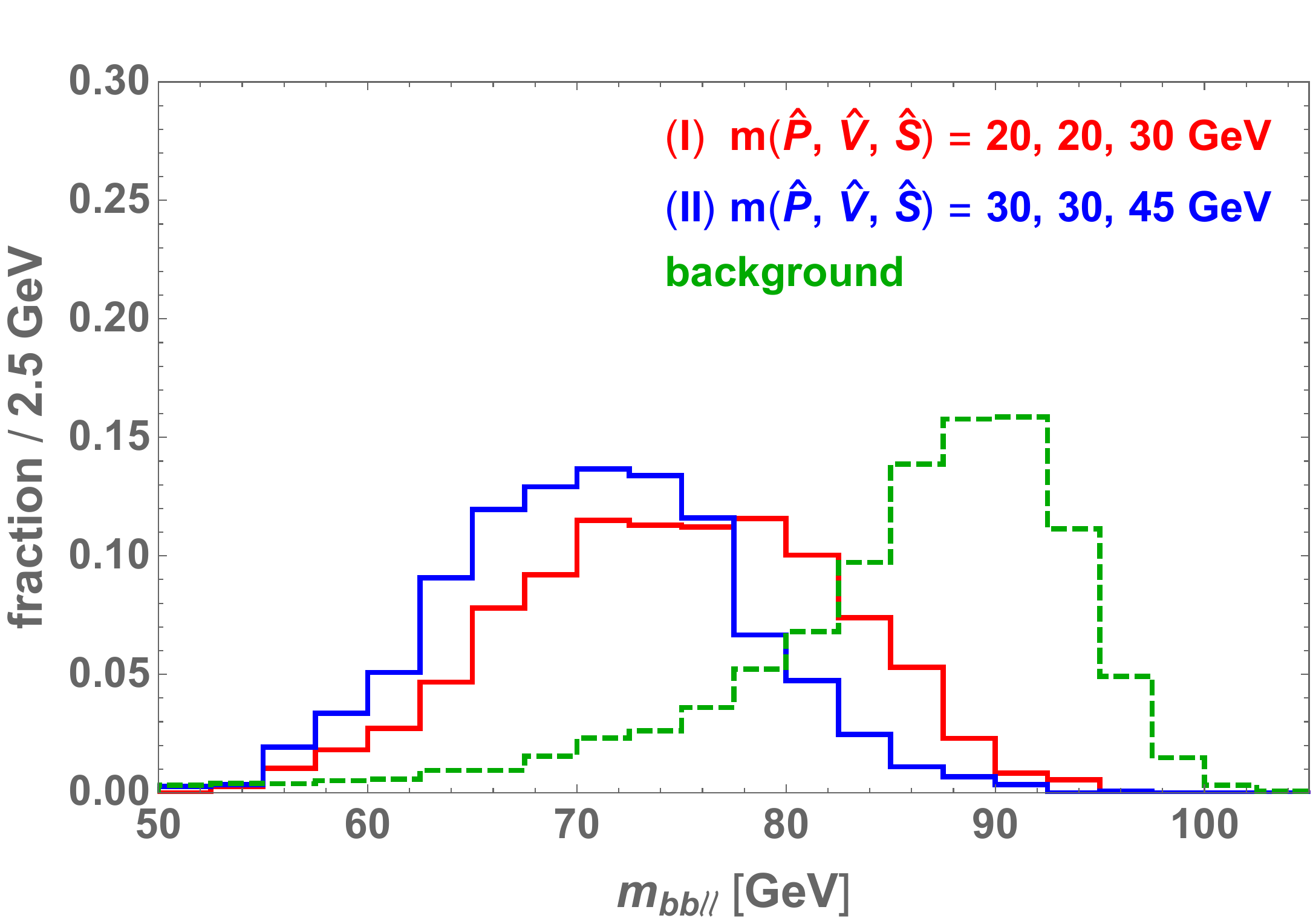}
\caption{\label{f.bbll_Zfact}Normalized distributions of the $b\bar{b}$ invariant mass ({\it left}) and total invariant mass of the $b\bar{b}\ell\ell$ system ({\it right}) for the $Z\to \omp \opp \to b\bar{b} \ell \ell + X$ signal and SM background at a $Z$ factory, after the basic selection of Eq.~\eqref{eq:Zfact_bbll_cuts}.} 
\end{figure}

\noindent Turning to the prospects at future $Z$ factories, we analyze first the $ee\to Z \to ( \omp \to b\bar{b} )(\opp  \to \lmm f\bar{f} \to \ell \ell f\bar{f})$ final state, where $\ell$ includes both electrons and muons. The main SM background is $ee \to b\bar{b}\ell\ell$, with amplitude at $O(g_w^4)$. We generate both signal and background using MadGraph5, interfaced with Pythia8 for parton showering. Detector simulation is performed with Delphes3, using the CEPC card (with, in particular, a $b$-tagging efficiency $\epsilon_b = 0.8$) but lowering the jet $p_T$ threshold to $5$ GeV and applying the same jet energy scale that we used for the LHC.\footnote{Namely, $\big( \frac{(2.5 - 0.15\, | \eta | )^2}{ p_T / \mathrm{GeV}} +1 \big)^{1/2}\,$. The default CEPC card does not apply any jet energy scale.} After the basic selection
\begin{equation} \label{eq:Zfact_bbll_cuts}
2\, b\,\mbox{-}\mathrm{jets} \;\mathrm{with}\; E_b > 10\;\mathrm{GeV}, |\eta_b| < 2.3,  \qquad 2\, \ell \;\mathrm{with}\; E_\ell > 5\;\mathrm{GeV}, |\eta_\ell| < 2.3\,,
\end{equation}
we obtain the normalized distributions shown in Fig.~\ref{f.bbll_Zfact}. We then impose the further cuts
\begin{equation}
| m_{\ell \ell} - m_{\lmm}|  < 0.5\;\mathrm{GeV}, \qquad m_{bb} \in [m_{\omp} - 10\;\mathrm{GeV}, m_{\omp} + 5\;\mathrm{GeV}], \qquad m_{bb\ell \ell} < 85\;\mathrm{GeV}.
\end{equation}
The total acceptance times efficiency for the signal is $(\mathcal{A}\,\epsilon)_{\rm tot}^{\rm I, \,II} = 12\%, 11\%$, whereas the expected background yield at TeraZ is of $65$ and $49$ events, respectively. The resulting $95\%$ CL bounds (calculated using Poisson statistics) are, in terms of the branching ratio $\mathrm{BR}(Z\to \omp \opp \to b \bar{b} \ell \ell + X)$, $2.6 \times 10^{-8}\,(1.2\times 10^{-10})$ at GigaZ~(TeraZ) for benchmark I, and $2.9 \times 10^{-8}\,( 1.3\times 10^{-10})$ for benchmark II. We stress that these constraints are derived assuming that the hidden mesons decay promptly. This is satisfied in most of the parameter space that can be probed at GigaZ, where the bounds translate to
\begin{equation} \label{eq:Zfactory_bbll_bounds}
(\mathrm{I}) \quad M \gtrsim 5.4\;\mathrm{TeV} \left( \frac{ f_{\omp \opp} }{1}\right)^{1/4} , \qquad\quad (\mathrm{II}) \quad M \gtrsim 5.2\;\mathrm{TeV} \left( \frac{ f_{\omp \opp} }{1}\right)^{1/4}, \qquad (\mathrm{GigaZ})
\end{equation}
since in this $\Lambda$ range the pseudoscalar has a lifetime $\lesssim \mathrm{mm}$ for $M\lesssim 5\;\mathrm{TeV}$. For larger $M$ the $\omp$ becomes long-lived, requiring a change in experimental strategy at TeraZ (discussed in Subsec.~\ref{sec:displaced}). Nonetheless, the above bounds on the $Z$ branching ratio can be relevant to other scenarios, where the decays remain prompt even for very small rates.

Finally, a comment is in order about the potential sensitivity of LEP1 data to this final state. The 4 LEP experiments recorded a combined total of $1.7 \times 10^7$ events at the $Z$ pole~\cite{ALEPH:2005ab}, hence after accounting for $\mathrm{BR}(Z\to \nu \bar{\nu}) = 0.20$ we estimate that $\sim 2.2 \times 10^7$ $Z$ bosons were produced. This yields the number of $Z\to \omp \opp \to b\bar{b} \ell \ell + X$ events, $N_S \approx 28\, f_{\omp \opp}\, (2\;\mathrm{TeV} / M)^4 (\mathcal{A}\,\epsilon)_{\rm tot}$. Given the relatively low $b$-tagging efficiency at LEP1, \mbox{$\epsilon_b \sim 0.3$}~\cite{ALEPH:2005ab}, we expect $< 3$ events and therefore no constraint for $M = 2\;\mathrm{TeV}$.

\subsubsection*{Higgs decays}
\noindent We expect the dominant Higgs two-body decays to be $h \to \omp \omp, \lmm \lmm, \opp\opp$, since $h \to \lmm \omp, \lmm \opp$ violate $C$ and $h\to \omp \opp$ violates $CP$. Here we concentrate on $h\to \lmm \lmm$ and $h\to \omp\omp$, which appear most promising. We discuss, drawing from the existing experimental and theoretical literature, a few searches for these decays that provide useful points of comparison with our results for $Z$ decays presented above. Defining $f^h_{XY}\leq 1$ as the fraction of Higgs decays to the hidden sectors that yield the $XY$ final state, we take $ f^h_{\lmm\lmm} , f^h_{\omp\omp}\sim 0.3$ as benchmarks.

We begin with $h\to \lmm\lmm \to b\bar{b}\mu \mu$, for which we apply the CMS search of Ref.~\cite{Sirunyan:2018mot} that we used to estimate the reach in $Z\to \omp \opp$. The CMS result is a bound $\mathrm{BR}(h \to aa \to b\bar{b} \mu\mu) \lesssim 2\times 10^{-4}$, weakly dependent on $m_a$ in the range $[20, 62.5]$~GeV. Neglecting the difference in acceptance between pseudoscalar and vector, this translates into \mbox{$M \gtrsim 0.55\,\mathrm{TeV}\, (f^h_{\lmm\lmm} / 0.3)^{1/4} (\alpha_d / 0.24)^{1/2} (c_g /4)^{1/2}$}, where we use $\mathrm{BR}(\lmm \to b\bar{b}, \mu\mu) \approx 0.15, 0.034$ and take as reference the coupling strength corresponding to $\Lambda =10\;\mathrm{GeV}$. This current constraint on $M$ is weaker than the one we obtained with the extremely conservative analysis of $Z\to \omp \opp$, see Eq.~\eqref{eq:Z_bbmumu_bounds}.

A second channel we consider is $h\to \lmm\lmm \to \ell\ell \ell^\prime \ell^\prime$. Here we directly apply the results of the ATLAS analysis in Ref.~\cite{Aaboud:2018fvk}, which targeted (among others) the $h\to Z_d Z_d$ signal~\cite{Curtin:2014cca}, where $Z_d$ is a dark photon kinetically mixed with hypercharge. Using $36.1$ fb$^{-1}$ of data, the search set a constraint $\mathrm{BR}(h\to Z_d Z_d) \lesssim 10^{-4}$ for $m_{Z_d}$ in the range $[15, 50]~\mathrm{GeV}$. Taking into account the different branching ratios to leptons, $\mathrm{BR}(\lmm \to \ell \ell) /  \mathrm{BR}(Z_d \to \ell \ell) \approx 0.068 / 0.30$ when summed over $e$ and $\mu$, we find $M \gtrsim 0.97\,\mathrm{TeV}\, (f^h_{\lmm\lmm} / 0.3)^{1/4}(\alpha_d / 0.24)^{1/2} (c_g / 4)^{1/2}$. Comparing with Eqs.~\eqref{eq:Z_4muX_results} and \eqref{eq:Z_4mu_results}, this current bound on $M$ is significantly weaker than those from $Z\to 4\mu\,(+X)$. Due to the democratic decays of $\lmm$ to SM fermions, $h\to \lmm\lmm$ yields a variety of other final states, many of which were discussed in the extensive survey of Ref.~\cite{Curtin:2013fra}.

Looking ahead to future Higgs factories, a particularly appealing prospect is the possibility to probe $h\to \omp \omp \to 4b$, which has a relatively large branching ratio due to $\mathrm{BR}(\omp \to b\bar{b})\sim 0.9$, and for which the sensitivity at FCC-ee will reach down to $\mathrm{BR}(h \to (b\bar{b})(b\bar{b})) = 3 \times 10^{-4}$ for $m_{\omp}$ in the range $[20, 60]$~GeV~\cite{Liu:2016zki}. This corresponds to the bound $M \gtrsim 1.4\,\mathrm{TeV}\, (f^h_{\omp\omp} / 0.3)^{1/4}(\alpha_d / 0.24)^{1/2} (c_g /4)^{1/2}$, much weaker than what can be achieved from $Z\to \omp \opp$ even at GigaZ, see Eq.~\eqref{eq:Zfactory_bbll_bounds}. These results illustrate in a quantitative manner the superior sensitivity of $Z$ over $h$ decays in probing this region of parameters.

\subsection{Two-body displaced decays}\label{sec:displaced}

For $\Lambda \sim 10$~GeV but larger $M$, at least some of the hidden mesons become long-lived particles (LLPs) for collider purposes, requiring different experimental strategies. In this subsection we discuss searches for the long-lived mesons at ATLAS and CMS, as well as at future $Z$ factories. As $M$ is increased, the $\omp$ and $\opp$ become long-lived first (the latter especially if $\Delta m$ is even moderately smaller than $\Lambda$, see Fig.~\ref{f.mesonDecay}), whereas the $\lmm$ remains prompt. We exploit this feature to propose HL-LHC searches where a mostly-hadronically decaying LLP ($\omp$ or $\opp$) is produced in association with a $\lmm$, whose prompt decay to $\mu^+\mu^-$ allows for efficient triggering and suppresses the SM background to a negligible level. A similar search was proposed in Ref.~\cite{Blinov:2017dtk}, taking as benchmark the process $Z\to A^\prime h_D$, where the dark photon $A^\prime$ decays promptly to leptons and the dark Higgs $h_D$ is the LLP. Furthermore, ATLAS has recently published~\cite{Aaboud:2018arf} a search for $h\to Z Z_d$, where $Z\to \ell \ell$ and the LLP $Z_d$ decays in the hadronic calorimeter (HCAL). 

\vspace{0.5cm}
In addition, we extend our analysis to TeraZ, where very large scales $M > 10\;\mathrm{TeV}$ can be probed. In this region of parameters the $\lmm$, too, can decay at a macroscopic distance from the interaction point. However, the trigger does not pose a problem and we expect that the combination of a (possibly) displaced dilepton pair from $\lmm \to \ell \ell$ with a displaced vertex (DV) from $\omp$ or $\opp$ will remove any SM backgrounds. As a consequence, when going from HL-LHC to TeraZ we do not make dramatic changes to our method for deriving projections.

\vspace{0.5cm}
A general difficulty is that the meson $c\tau$ and $\mathrm{BR}(Z, h \to \overline{\psi} \psi)$ have inverse scaling with $M$, so that macroscopic decay lengths necessarily correspond to very small branching ratios to the hidden sectors. Nevertheless, the very large statistics that will be collected at the HL-LHC and TeraZ and the absence of backgrounds result in promising sensitivity.

\subsubsection*{Displaced pseudoscalar decays}\label{sssec:Zmumubb}
\noindent Combining Eqs.~(\ref{eq:ZBr}),(\ref{eq:hBr}), and (\ref{eq:tau_P}) we can write 
\begin{eqnarray}
c\tau_{\omp}  \sim 1 \;\mathrm{cm }\, \left( \frac{m_{\omp}}{20\;\mathrm{GeV}} \right)^2 \left( \frac{10\;\mathrm{GeV}}{\Lambda} \right)^5 \left( \frac{\Lambda}{\constmass} \right)^2 &\times&
\Bigg\{ \frac{8.3 \times 10^{-8}}{\mathrm{BR}(Z\to \overline{\psi}_{B,C} \psi_{B,C})}
 \\
&\mbox{or}& \;\;\; \frac{1.3 \times 10^{-6}}{\mathrm{BR}(h\to \overline{\psi}_{B,C} \psi_{B,C}) } \left( \frac{\alpha_d}{0.24} \right)^2  \left( \frac{c_g}{4} \right)^2
 \Bigg\}\, , \nonumber
\end{eqnarray}
where the dependence on $M$ cancels out. For reference, the HL-LHC (running at $\sqrt{s} = 14$~TeV) will produce approximately $1.8\times 10^{11}$ $Z$ and $1.6\times 10^8$ Higgs bosons, hence we concentrate on $Z$ decays and specifically on $Z \to \omp \opp$, which yields a $\omp \to f\bar{f}$ DV (where about $90\%$ of the time $f = b$, and otherwise $f = \tau, c$), together with $\opp \to \lmm f\bar{f}$. The decays of both $\opp$ and $\lmm$ are assumed to be prompt. We select $\lmm \to \mu\mu$, performing a parton-level MadGraph5 simulation. We require the muon pair to pass the dimuon trigger requirements:  $p_T^\mu > 17, 8$~GeV and $| \eta_\mu | < 2.5$~\cite{Sirunyan:2018mot}.\footnote{We have checked that because of the higher transverse momentum thresholds for the dielectron trigger, $p_T^e > 23, 12$~GeV~\cite{Sirunyan:2017lae}, the contribution of $\lmm \to ee$ would be relatively suppressed by about one order of magnitude, so we neglect it.} 

\begin{figure}
\includegraphics[width=0.49\textwidth]{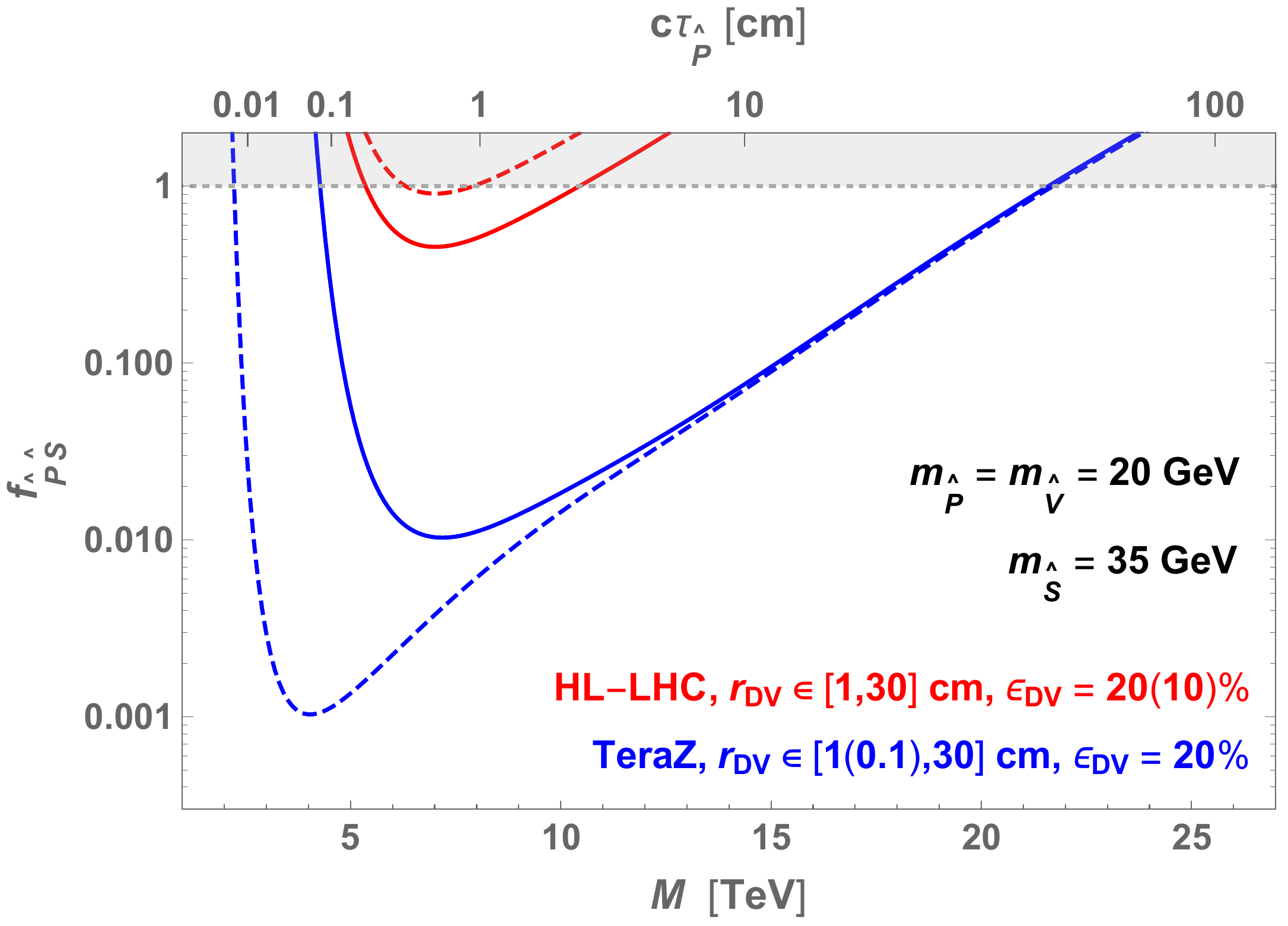}\hspace{1mm}
\includegraphics[width=0.49\textwidth]{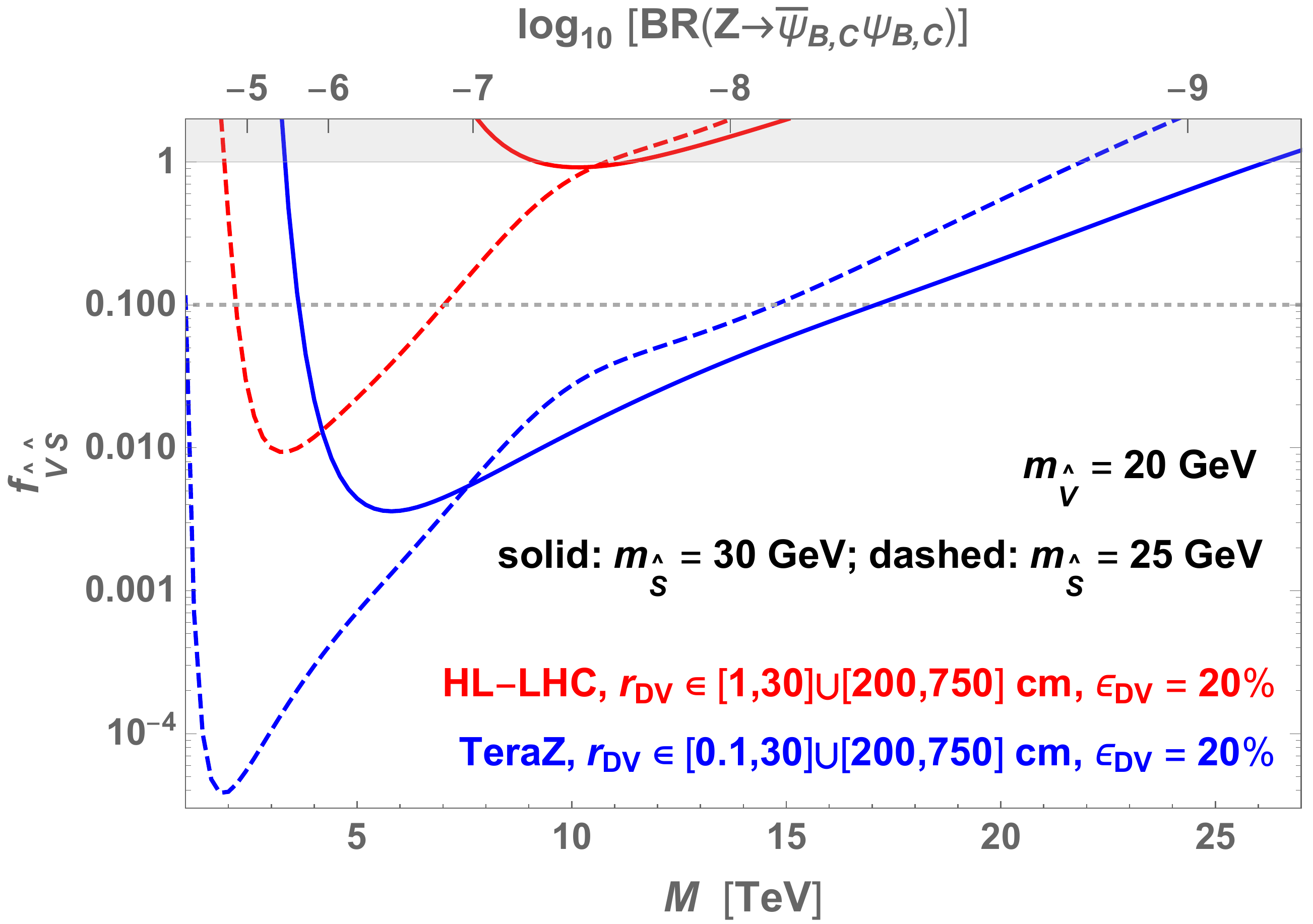}
\caption{\label{f.displaced_new}{\it Left:} projected bounds on $f_{ \omp\opp}$ from future searches for $Z\to \omp \opp$, where $\omp$ is long-lived while $\opp \to (\lmm \to \ell\ell) f\bar{f}$. {\it Right:} projected bounds on $f_{\lmm \opp}$ from future searches for $Z \to \lmm \opp$, where $\opp$ is long lived while $\lmm \to \ell\ell$. All bounds are at $95\%$ CL, assuming negligible SM background. We take $\ell = \mu~(e$ or $\mu)$ at the LHC~(TeraZ), and the dilepton pair must be prompt at the LHC, but can be displaced at TeraZ. The dotted gray lines correspond to the educated guesses $f_{\omp \opp,\, \lmm \opp} \sim 1, 0.1$ discussed in Subsec.~\ref{sec:prompt}.} 
\end{figure}

Furthermore, we impose $| \eta_{\omp} | < 2.5$ and for each surviving event we integrate the $\omp$ decay probability distribution, determined by the four-momentum of $\omp$ and its proper lifetime in Eq.~\eqref{eq:tau_P}, over the volume of the detector where the DV can be reconstructed. The latter is taken to be an annulus with radii $r \in [1, 30]$~cm, approximately corresponding to the capability of the ATLAS inner tracking detector~(ID)~\cite{Aad:2015uaa}, with efficiency for hadronic DV reconstruction equal to a constant $\epsilon_{\rm DV}$. We take $\epsilon_{\rm DV} = 20\,(10)\%$ as an optimistic~(conservative) benchmark, and by requiring $3$ signal events -- corresponding to a $95\%$ CL exclusion if the background is negligible -- we derive the solid~(dashed) red curve in the left panel of Fig.~\ref{f.displaced_new}. Note that for $M \sim 11$~TeV, the largest scale accessible at the HL-LHC, the chosen benchmark spectrum gives $c\tau_{\opp} \sim 1$~mm and $c\tau_{\lmm} \sim 0.6$~mm, hence the dimuon pair can still be considered prompt.

The TeraZ analysis proceeds along similar lines, but we include $\lmm$ decays to both electrons and muons, requiring $E_\ell > 10\;\mathrm{GeV}$ and $| \eta_{\ell} | < 2.3$, as well as $| \eta_{\omp} | < 2.3$. For the DV coverage we consider two options, one identical to the LHC to facilitate the comparison, and one where efficient hadronic DV reconstruction is extended down to $r = 1\,\mathrm{mm}$, which is expected to be easily achievable in the absence of pileup (see e.g. Ref.~\cite{BeachamTalk2018} for a recent discussion). In the left panel of Fig.~\ref{f.displaced_new}, these two scenarios are shown in solid blue and dashed blue, respectively. Both assume a constant efficiency $\epsilon_{\rm DV} = 20\%$, and negligible background. 

\subsubsection*{Displaced scalar decays}
\noindent Due to the $\sim (\Delta m)^7$ dependence of Eq.~\eqref{eq:Slifetime}, the lifetime of the scalar meson $\opp$ is very sensitive to the mass splitting with the $\lmm$, and even a mild hierarchy $\Delta m < \Lambda$ results in macroscopic decay lengths for $M \gtrsim \mathrm{few}$~TeV. To probe this LLP at the HL-LHC we choose the $Z\to \lmm \opp$ decay, selecting $\lmm \to \mu\mu$ which is assumed to be prompt. The mostly-hadronic DV\footnote{When calculating the expected signal rate we subtract the fraction of events where $\opp\to (\lmm \to \nu\bar{\nu})f\bar{f}$, in which case the tracks resulting from $f\bar{f}$ alone are likely too soft for DV reconstruction.} from $\opp \to \lmm f\bar{f}$ (or $\opp \to f\bar{f}$) is reconstructed either in the ID or in the HCAL$\,+\,$muon spectrometer volume, taken to be $r\in [2,7.5]$ m~\cite{Aad:2015uaa}. The inclusion of the outer detector is important due to the long lifetime for small $\Delta m$: for example, for $\Delta m = \Lambda/2 = 5\;\mathrm{GeV}$ we find $c\tau_{\opp} \sim 60$~cm at $M = 10$~TeV. The selection requirements on the muons are the same as in the analysis of displaced $\omp$, and we assume a constant $\epsilon_{\rm DV} = 20\%$ across the whole detector volume. At TeraZ we include both $\lmm \to ee, \mu\mu$ and assume a similar DV coverage, except for the already-mentioned extension down to $r = 1$~mm. The resulting sensitivities on $f_{\lmm \opp}$ are shown in the right panel of Fig.~\ref{f.displaced_new}, for two different values of $\Delta m$. The $\Delta m = \Lambda$ scenario (solid curves) is qualitatively similar to the case of displaced $\omp$ decays, whereas for $\Delta m = \Lambda/2$ (dashed curves) the scalar is long-lived already at $M \sim \mathrm{few}$~TeV, corresponding to larger $Z$ decay rates to the hidden sectors. This results in a better reach on $f_{\lmm \opp}\,$.

Finally, for the vector meson we find, combining Eqs.~(\ref{eq:ZBr}),$\,$(\ref{eq:hBr}) and (\ref{eq:tau_V}),
\begin{eqnarray}
c\tau_{\lmm}  \sim 1 \;\mathrm{cm }\, \left( \frac{20\;\mathrm{GeV}}{m_{\lmm}} \right)^2 \left( \frac{10\;\mathrm{GeV}}{\Lambda} \right)^3 &\times&
\Bigg\{ \frac{1.5 \times 10^{-9}}{\mathrm{BR}(Z\to \overline{\psi}_{B,C} \psi_{B,C})} 
 \\
&\mbox{or}& \;\;\; \frac{2.5 \times 10^{-8}}{\mathrm{BR}(h\to \overline{\psi}_{B,C} \psi_{B,C})} \left( \frac{\alpha_d}{0.24}\right)^2 \left( \frac{c_g}{4} \right)^2 
\Bigg\} ,  \nonumber
\end{eqnarray}
corresponding to $\sim 300$ $Z$ events and $\sim 4$ $h$ events after $3$ ab$^{-1}$ at $14$~TeV for the reference parameters. Clearly, observing displaced $\lmm$ decays at the HL-LHC will be extremely challenging. The prospects may be better at TeraZ, thanks in particular to the likely improved DV reconstruction efficiency at small displacements.

\subsection{Decays to hidden jets}\label{sec:emerging}

For sufficiently small confinement scale, roughly $\Lambda \lesssim 2$-$3\;\mathrm{GeV}$, $Z$ and Higgs decays to the hidden sector result in parton showers, producing jets of hidden mesons. Unless $M$ is very low, the mesons are long-lived (see Fig.~\ref{f.mesonDecay}) and therefore the hidden jets include a significant fraction of displaced vertices, realizing emerging jet-like phenomenology~\cite{Schwaller:2015gea}.\footnote{The related semivisible jets~\cite{Cohen:2015toa} occur when some of the hidden mesons decay promptly while others are stable on collider timescales. The discussion of Sec.~\ref{sec:light_resonances} shows that this is unlikely to happen in our setup.} As the vector meson $\lmm$ decays democratically to SM fermions, LHCb is especially well suited to probe our type of emerging jets, by resolving the decay of a single $\lmm \to \mu \mu$ inside the jet cone with the Vertex Locator (VELO). The sensitivity is maximal for lifetimes of $O(\mathrm{cm})$. On the other hand, tracker-based searches for emerging jets at CMS~\cite{Sirunyan:2018njd} and ATLAS must rely on hard cuts that suppress our signals to a negligible level. Therefore we need to require the production of an associated object, for example $pp \to ZV$ with $V$ an EW gauge boson that decays leptonically, or $pp\to Zj$ with a hard initial-state radiation jet. This ensures efficient triggering, but significantly reduces the signal rate while the backgrounds remain appreciable, and we find that these searches cannot compete with the LHCb sensitivity.\footnote{Emerging jet searches at ATLAS and CMS can, however, be important in scenarios where light hidden pions dominate the hadronization in the hidden sector~\cite{DarkPions}.}\enlargethispage{-0.5cm}

\subsubsection*{Reach at LHCb}
\noindent In $Z \to 2$ hidden jets events, the VELO can detect a single $\lmm\to\mu\mu$ DV within one of the jets~\cite{Pierce:2017taw}. This requires the vector meson to have transverse decay length between $6$ and $22$~mm~\cite{Ilten:2016tkc} and pseudorapidity within the LHCb coverage, $\eta \in [2,5]$. We simulate $Z$ decay events in Pythia8, choosing the benchmark $m_{\omp} = m_{\lmm} = 1$~GeV and $\Lambda= \mu_\psi = 0.5$~GeV. We impose that the average numbers of mesons in each jet satisfy $\left\langle N_{\omp} \right\rangle / \left\langle N_{\lmm} \right\rangle = 1/3\,$, as expected from a counting of the spin degrees of freedom, and for simplicity we neglect heavier hadrons in the shower, including the $\opp$ and other mesons, as well as the baryons. We expect the baryons to make up a small, invisible component of the jets, thus only slightly reducing the signal rate. We find that the simulation produces an average of $\sim7$ hidden mesons in a jet. Each $\lmm\to \mu\mu$ DV with $p_T^{\lmm}>1\;\mathrm{GeV}$, transverse decay length between $6\,$-$22$~mm and $\eta\in [2,5]$ is assumed to be reconstructed with constant efficiency $\epsilon_{\mu\mu} = 0.5\,$. The efficiency degradation due to the overlap with other nearby tracks is not explicitly mentioned in preceding studies~\cite{Ilten:2016tkc,Pierce:2017taw}, but it is likely to reduce $\epsilon_{\mu\mu}$ significantly. Therefore, we reject non-isolated events where the $\mu\mu$ vertex is accompanied by one or more $\lmm$ visible decays with transverse decay length shorter than $22$~mm and $\Delta R<0.4$. Because $\left\langle N_{\omp} \right\rangle$ is small and the $\omp$ decay length is much longer than that of $\lmm$, we neglect the effects of $\omp$ decays. 
\begin{figure}[t]
\includegraphics[scale=0.5]{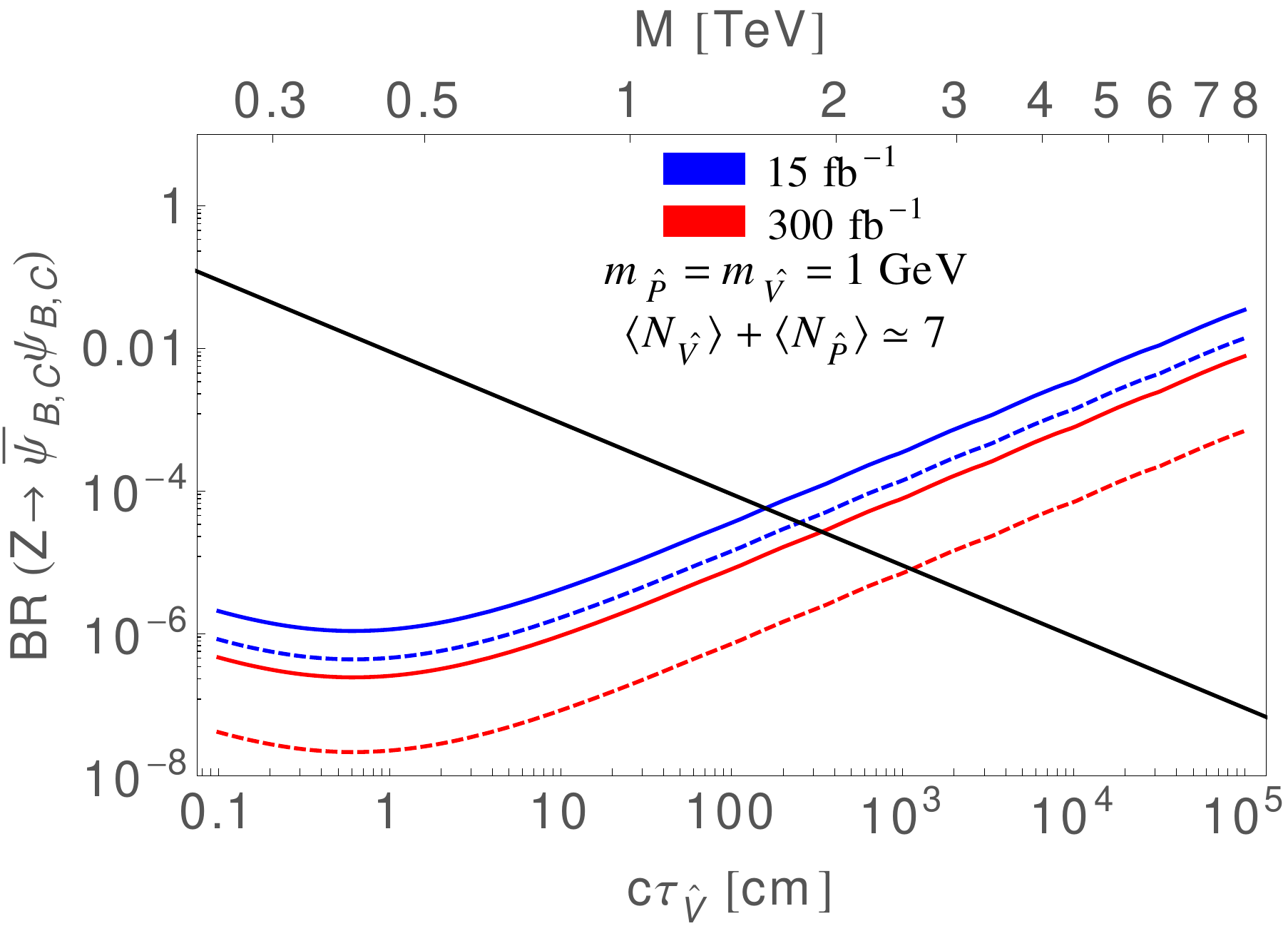}
\caption{\label{f.eJet}Projected limits on $\mathrm{BR}(Z\to \overline{\psi}_{B,C} \psi_{B,C})$ from the search for single $\lmm \to \mu\mu$ DV inside an emerging jet at LHCb. Solid~(dashed) lines correspond to the standard background count~(a background-free scenario). The black line corresponds to the theoretical prediction, $\mathrm{BR} \approx 9 \times 10^{-3} / (c\tau_{\lmm}/\mathrm{cm})$, as derived from Eqs.~\eqref{eq:ZBr} and \eqref{eq:UpsilonWidth}.}
\end{figure}

The SM background to such displaced $\mu\mu$ vertices is expected to be about 25 events for 15~fb$^{-1}$ of integrated luminosity~\cite{Ilten:2016tkc}. To estimate the ultimate sensitivity achievable at LHCb, we also compute the constraints by assuming negligible SM background. The results are shown in Fig.~\ref{f.eJet}, where the solid (dashed) curves correspond to standard (negligible) background. Due to the moderate typical boost factor, the sensitivity is optimal for \mbox{$c\tau_{\lmm} \simeq 1$~cm}, where it reaches $Z$ branching ratios down to $O(10^{-7})$ at the HL-LHC. In our setup, the bounds translate to $M\gtrsim 1.6\,(2.0)$~TeV for $L=15\,(300)$~fb$^{-1}$ assuming the standard background count, while in the background-free case we find $M\gtrsim 1.8 \,(2.7)$~TeV.

At future $Z$ factories, searches for emerging jet signals will greatly benefit from the straightforward triggering. The corresponding analysis is described in Appendix~\ref{App:EMJ_Zfactories}, where we find that at TeraZ the sensitivity will reach $\mathrm{BR}(Z \to \overline{\psi}_{B,C} \psi_{B,C}) \sim O(10^{-8})$ for \mbox{$c\tau_{\lmm} \sim O(1$-$10)$~cm}. As Fig.~\ref{f.eJet} shows, in our setup this will mostly probe parameter space already accessible at LHCb. In more elaborate scenarios, however, the $Z$ branching ratio to hidden fermions and the vector meson lifetime may be decoupled, for example by extending the model to allow additional decay channels for $\lmm$. In this case, a future $Z$ factory could provide the crucial test.

\subsection{Vector meson as dark photon}\label{sec:Bmeson}
The vector meson $\lmm$ couples to the SM through the $Z$ boson, with a coupling structure similar to that of a dark photon $\gamma_D$. For a broad class of dark photons, the interactions with the SM can be written as
\begin{equation} 
\mathcal{L}=-A_D^\mu\left(\varepsilon\, e J_\mu^\text{EM}+\varepsilon_Z\frac{g_Z}{2}J_\mu^\text{NC} \right),\label{e.kmix}
\end{equation}
where $A_D$ is the dark photon field and $J^\text{EM}$ and $J^\text{NC}$ are the electromagnetic and weak-neutral currents, respectively.\footnote{The currents are defined as $J_\mu^\text{EM} = \sum_f Q_f \overline{f}\gamma_\mu f$ and $J_\mu^\text{NC} = \sum_f \overline{f} \gamma_\mu (v_f - a_f \gamma_5)f$.} If a single source of kinetic mixing generates both of the operators in Eq.~\eqref{e.kmix}, then the coupling to $J^{\rm EM}$ is the more sensitive probe of the interaction. This is the case that is most frequently studied. However, in our scenario the constituent fermions do not couple to the photon at tree level and only the second operator in Eq.~\eqref{e.kmix} is relevant, which can be seen as originating from an effective $Z$-$\lmm$ mass mixing term $\varepsilon_Z m_Z^2 Z_\mu \lmm^\mu$~\cite{Davoudiasl:2012ag}. This coupling can lead to a variety of signals in low-energy experiments, such as parity violation tests and flavor-changing neutral current (FCNC) meson decays, in particular $B\to K \gamma_D$ and $K\to\pi\gamma_D$~\cite{Davoudiasl:2012ag,Dror:2018wfl}. 

By matching to Eq.~\eqref{eq:UpsilonWidth} we identify
\begin{equation} \label{eq:epsilonZ}
\varepsilon_Z \simeq g_Z \sqrt{\frac{N_d}{2}} \frac{m_t^2}{M^2}  \frac{|\psi(0)| m_{\lmm}^{1/2}}{m^2_Z}\,\approx\, 3.2 \times 10^{-7}\left(\frac{\Lambda}{1\,\text{GeV}} \right)^{3/2}\left( \frac{m_{\lmm}}{2\,\text{GeV}}\right)^{1/2}\left(\frac{2\,\text{TeV}}{M} \right)^2,
\end{equation} 
where we have assumed $m_{\lmm} \ll m_Z$ and included an extra factor of $\sqrt{2}$ to account for the two hidden sectors $B,C$.\footnote{A small coupling to the electromagnetic current is induced by the magnetic dipole operator \mbox{$\sim \alpha_W   m_t^2 m_\psi \overline{\psi} \sigma^{\mu \nu} \psi \,e F_{\mu \nu} / (\pi M^4)$}, generated at one loop through the $W$ boson and the electrically-charged heavy fermion $\Psi^-$. Neglecting $O(1)$ factors, we estimate $\varepsilon / \varepsilon_Z \sim \alpha_W  e \,m_Z^2 / ( \pi g_Z M^2) \approx 10^{-5}\, (2\;\mathrm{TeV}/M)^2$.} Very recently, Ref.~\cite{Dror:2018wfl} derived strong FCNC decay bounds on $\varepsilon_Z$ for a generic light vector $X_\mu$. In our setup, though, the NP contribution to a given SM final state should be smaller than the corresponding perturbative rate, $\mathrm{BR}(B \to K f\bar{f})_{\rm NP} \lesssim \mathrm{BR}(B \to K \psi \overline{\psi})$, where $f$ is a SM fermion and we have taken $B$ decays as example. This allows us to conservatively estimate
\begin{equation} \label{eq:FCNC_mesonDecays}
\frac{\mathrm{BR}(B \to K f\overline{f})_{\rm NP}}{\mathrm{BR}(B \to K f\overline{f})_{\rm SM}} \lesssim \left( \frac{m_t}{M} \right)^4 \frac{\mathrm{BR}(\hat{V} \to \nu \overline{\nu})}{\mathrm{BR}(\hat{V} \to f \overline{f})} \,,
\end{equation}
where we assumed dominance of the $Z$ penguin over the box amplitudes and used Eq.~\eqref{eq:Zpsipsi_coupling}. An analogous expression applies to $K$ decays. For any $f$, the RHS of Eq.~\eqref{eq:FCNC_mesonDecays} is below a percent if $M > 1\;\mathrm{TeV}$, showing that the precision needed to probe the NP is well beyond the current one~\cite{Dror:2017nsg}.\footnote{Reference~\cite{Dror:2018wfl} exploited the emission of the longitudinal mode of $X_\mu$, which can lead to $\sim (m_{\rm EW} / m_X)^2$ enhancement. However, our vector meson $\lmm$ is a composite particle, which ``dissolves'' into its constituents at energies not far above its mass, so such enhancement does not apply.} We do not expect non-perturbative corrections to change this conclusion.\enlargethispage{-1cm} 

Looking ahead, it is interesting to ask whether future dark photon searches~\cite{Bauer:2018onh,Beacham:2019nyx} will be able to probe the $\lmm$. In general, dark photon production is simply accomplished by taking a visible photon production channel and replacing one visible photon with a dark photon. For instance, a fixed-target setup with target nucleus $N$ and beam particle $e$ produces bremsstrahlung, $eN\to eN\gamma$. Consequently, dark photons $\gamma_D$ can be produced through $eN\to eN\gamma_D$. Similarly, lepton colliders produce a visible photon recoiling off a dark photon through $e^+e^-\to\gamma\gamma_D\,$, and hadron colliders exploit Drell-Yan production $q\overline{q}\to\gamma_D$. In addition, as discussed above, the dark photon can be produced in meson decays if kinematically allowed. Comprehensive reviews can be found, for example, in Refs.~\cite{Essig:2013lka,Alexander:2016aln}. From Eq.~\eqref{e.kmix} we read that in order to gain a first, crude impression of the future reach, we can treat the $\lmm$ as a kinetically mixed dark photon with ``effective coupling''
\begin{equation}
\varepsilon_\text{eff} \sim \frac{1}{2} \sqrt{\frac{\alpha_Z}{\alpha}} \,\varepsilon_Z \,\approx\, 3.8 \times 10^{-7} \left(\frac{\Lambda}{1\,\text{GeV}} \right)^{3/2}\left( \frac{m_{\lmm}}{2\,\text{GeV}}\right)^{1/2}\left(\frac{2\,\text{TeV}}{M} \right)^2 ,
\end{equation}
where we have neglected the differences between the electric and weak charges of the SM fermions. However, the $\lmm$ has an appreciable decay width to the SM neutrinos, since it mixes with the $Z$ rather than the photon. This reduces the $\lmm$ rates to visible particles, especially for smaller $m_{\lmm}$, where decays to light quarks are cut off by the meson masses. To account for this we rescale $\varepsilon_\text{eff}$ by $\text{BR}(\lmm\to\text{visible})^{1/2}$ as a function of $m_{\lmm}$.
\begin{figure}
\includegraphics[width=0.6\textwidth]{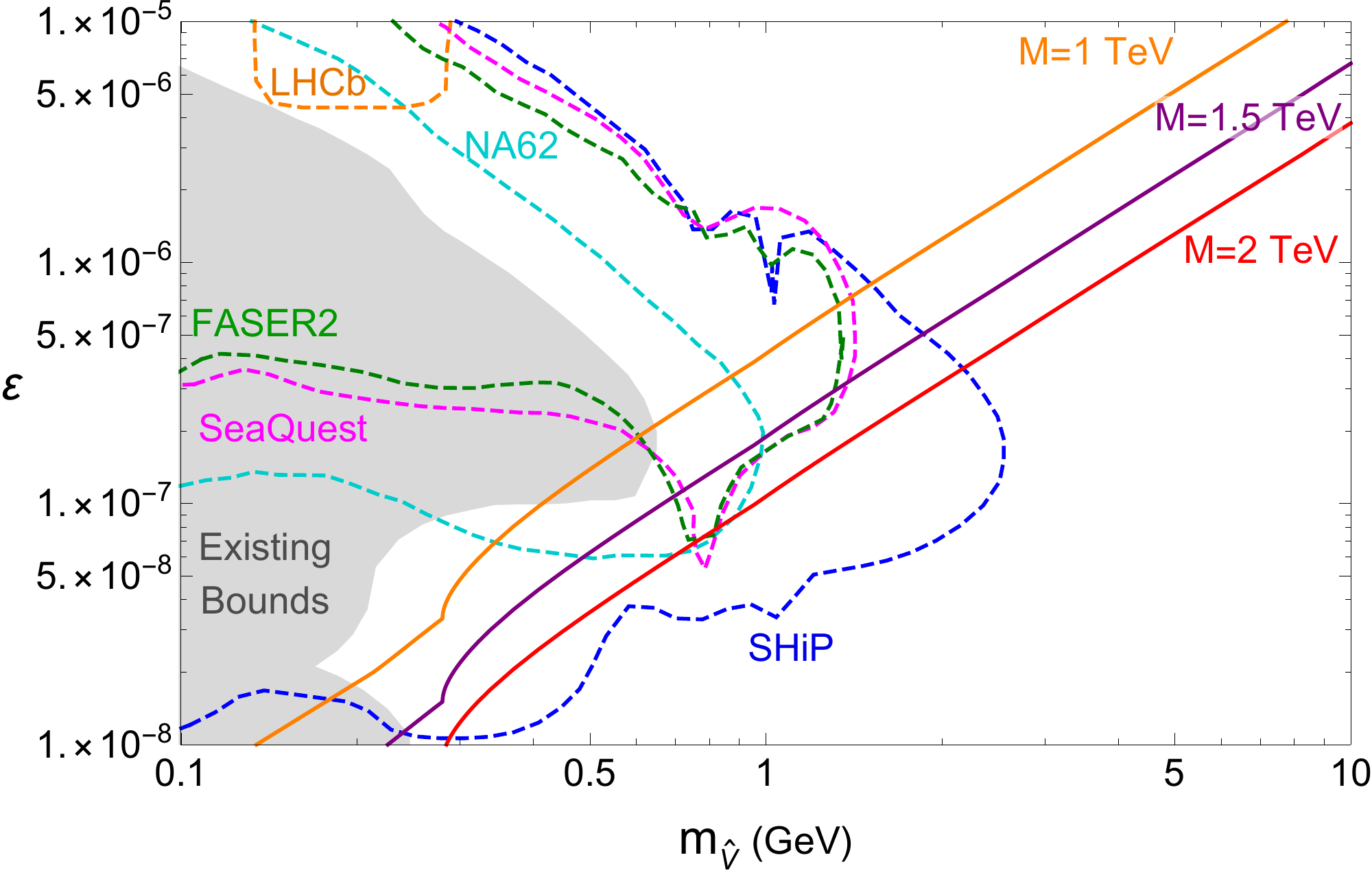} 
\caption{\label{f.darkPhoton} Illustration of where the $\lmm$ lies in the dark photon parameter space (solid lines), for three $M$ benchmarks. Also shown are current exclusions (gray-shaded region) and the projected reach of future experimental probes (dashed lines).}
\end{figure}

In Fig.~\ref{f.darkPhoton} we show where $\lmm$ falls in the standard dark photon parameter space. The solid lines trace out the relationship between the vector mass and $\varepsilon_{\rm eff}$ for three $M$ benchmarks, assuming $m_{\lmm} = 2 \Lambda$. The gray-shaded region indicates the existing bounds on dark photons, as collected in Ref.~\cite{Beacham:2019nyx}. The dashed lines denote projected limits from a collection of future experiments: we expect that FASER2~\cite{Ariga:2018uku}, SeaQuest~\cite{Berlin:2018pwi} and NA62~\cite{Beacham:2019nyx}, as well as ultimately SHiP~\cite{Alekhin:2015byh}, will be sensitive to hidden vector mesons with masses below a few GeV. These preliminary results provide solid motivation for a more detailed analysis of this physics, which we leave as an interesting direction for future work.

\subsection{Invisible decays}\label{sec:invisible}
If the hidden mesons are very long-lived, as is expected at small $\Lambda$, then invisible $Z$ and Higgs decays constitute the main search strategy. Currently, the strongest constraint comes from the LEP1 measurement of the $Z$ invisible width, requiring $\Delta \Gamma_Z^{\rm inv} < 2\;\mathrm{MeV}$ at $95\%$ CL~\cite{ALEPH:2005ab}. Using Eq.~\eqref{ZFermions}, this translates to $M > 0.8\;\mathrm{TeV}$ for $m_\psi =0\,$. From Eq.~\eqref{eq:hBr} we obtain then the current upper bound on the invisible Higgs branching ratio, $\mathrm{BR}(h \to \hat{g}_{B,C}\hat{g}_{B,C}) \lesssim 1.4 \times 10^{-2}\, (\alpha_d / 0.24)^2 (c_g / 4)^2$, which is likely out of reach at the LHC even in the high-luminosity phase~\cite{ATL-PHYS-PUB-2013-014}. 

Turning to future colliders, an FCC-ee run in TeraZ mode will be able to improve the precision on the number of light neutrino species by a factor $7$~\cite{Abada:2019lih}. This roughly corresponds to a reduction by the same factor of the uncertainty on $\Gamma_Z^{\rm inv}$, leading to a $95\%$ CL projected constraint $\Delta \Gamma_Z^{\rm inv} \lesssim 0.4\;\mathrm{MeV}$,\footnote{We have assumed the future measurement will agree with the SM prediction. Recall that at LEP1 $\Gamma_{Z}^{\rm inv}$ was measured $\lesssim 2\,\sigma$ below the SM~\cite{ALEPH:2005ab}, so the current bound on $\Delta \Gamma_Z^{\rm inv}$ is stronger than the expectation.} or $M \gtrsim 1.2\;\mathrm{TeV}$. A more careful analysis including correlations would affect this estimate only mildly. This method is severely limited by the systematic uncertainty affecting the measurement of the integrated luminosity~\cite{Abada:2019lih}, and in fact the ultimate precision can already be achieved at GigaZ. Runs at higher center-of-mass energy will be able to constrain $\Delta \Gamma_Z^{\rm inv}$ through radiative return $e^+ e^- \to Z\gamma$ events, although current estimates suggest an improvement of only $20\%$ on the uncertainty with respect to the measurement at the $Z$ pole~\cite{Abada:2019lih}. 

Finally, we should mention that FCC-hh running at $100$~TeV will be able to achieve an impressive limit on the Higgs branching ratio to new invisible particles, $\mathrm{BR} (h \to \mathrm{invisible}) < 2.5 \times 10^{-4}$ at $95\%$ CL including systematic uncertainties~\cite{L.Borgonovi:2642471}, by exploiting the large sample of Higgses produced with large transverse momentum. Using Eq.~\eqref{eq:hBr}, we translate this to $M > 2.2\,\mathrm{TeV}\,  (\alpha_d / 0.24)^{1/2} (c_g / 4)^{1/2}$.

\section{Conclusions\label{sec:conclusions}}
Motivated by a possible realization of the tripled top model for neutral naturalness, we have studied a confining hidden sector where one fermion $\psi$ is light compared to the confinement scale $\Lambda$. The latter is taken in the range $0.1\;\mathrm{GeV} \lesssim \Lambda \lesssim 15\;\mathrm{GeV}$, as generally motivated by $2$-loop naturalness considerations. The couplings of the hidden sector fields to the SM are mediated by EW-charged particles with TeV-scale mass $M$, and are described at low energies by a handful of dimension-$6$ operators. These determine both the production and the decays of the hidden sector mesons through the $Z$ and Higgs portals. Since the theory does not possess light pNGBs, several of the lightest mesons are important for phenomenology, resulting in a complex pattern of signatures. We performed a survey of these, identifying several regions of the parameter space -- which is characterized primarily by $\Lambda$ and $M$ -- with distinctive phenomenological properties. We have found that the $Z$ portal, which was not considered in the previous hidden valley literature, has dramatic implications for the prospects to detect the hidden sector. In particular, the enormous numbers of $Z$ events that will be collected at the HL-LHC and at a future $Z$ factory, allow for unprecedented sensitivity. 

For large $\Lambda \sim O(10) \;\mathrm{GeV}$, the $Z$ decays to two-body hidden meson final states, followed by prompt decays back to the SM. We showed that adaptations or extensions of current LHC searches can probe $M$ up to $\sim 3\;\mathrm{TeV}$ in the high-luminosity phase, whereas a future GigaZ run at the $Z$ pole will be able to reach $M \sim 5\;\mathrm{TeV}$. For even larger $M$ the hidden mesons become long-lived, and we have found that a TeraZ run will probe scales up to $M\sim 20 \;\mathrm{TeV}$ for negligible SM background. These results demonstrate in a quantitative way the power of a future $Z$ factory to directly probe the confining hidden sector. For smaller $\Lambda \sim O(1)\;\mathrm{GeV}$ the $Z$ decays produce jets of hidden mesons, which are typically long-lived. At the LHC the best sensitivity to this scenario is obtained at LHCb, by resolving single displaced $\lmm \to \mu\mu$ decays within the jets, where $\lmm$ is the hidden vector meson. In addition, since the $\lmm$ couples to the SM through mixing with the $Z$, it can be probed by an array of planned experiments that will search for dark photons at the intensity frontier. 

Our work can be extended in a number of ways. First of all, we stress that our phenomenological study is intended to be only an initial survey of the many possibilities available, and several areas deserve to be analyzed in greater detail. At the LHC, searches for $Z$ decays to mesons with masses in the $10$-$40\;\mathrm{GeV}$ range strongly benefit from keeping the selection cuts as soft as possible, and we encourage ATLAS and CMS to extend their analyses in that direction. Another topic that warrants further attention is the future experimental sensitivity to light hidden vectors mixed with the $Z$ boson (such as our $\lmm$), which we have only sketched briefly, but should be analyzed in a systematic way. In addition, in this paper we have focused our attention on the $Z$ portal to the hidden sector, which constitutes a main novelty of our setup, while Higgs decays have typically played a marginal role. This picture may change at the FCC-hh, however, thanks to the large sample of Higgs bosons that will be collected (approximately $3\times 10^{10}$ for $30$ ab$^{-1}$ at 100~TeV~\cite{L.Borgonovi:2642471}); the associated new physics reach certainly deserves to be investigated. Finally, our results reinforce the relevance for phenomenology of $1$-flavor (hidden) QCD, whose properties can only be reliably studied using non-perturbative methods. We believe this theory merits attention from the part of the lattice QCD community with interest in BSM physics, and further results beyond the partial ones of Ref.~\cite{Farchioni:2007dw} would be most welcome.

\vspace{8mm}

\noindent{\bf Acknowledgments} We thank J.L.~Feng, M.~Reece, Y.~Tsai and M.~Williams for useful conversations. ES is grateful to O.~Mattelaer for help with MadGraph5. H-CC and CBV are supported by Department of Energy Grant number DE-SC-0009999. LL is supported by the General Research Fund (GRF) under Grant No.~16312716, which was issued by the Research Grants Council of Hong Kong S.A.R. ES has been partially supported by the DFG Cluster of Excellence 2094 ``ORIGINS: From the Origin of the Universe to the First Building Blocks of Life,'' by the Collaborative Research Center SFB1258 and BMBF grant no. 05H18WOCA1, and thanks the MIAPP for hospitality during the completion of this work.

\appendix

\section{$T$ Parameter in the Tripled Top Framework} \label{app:Tparameter}
We present here the calculation of the $T$ parameter (precisely, of $\widehat{T} = \alpha T = \rho - 1$) in the tripled top framework. We first consider the model used as motivation for the present paper.  A similar calculation can be done for the original model of Ref.~\cite{Cheng:2018gvu}, which is also discussed. To begin with, there is a contribution from loops of the $A$ stop/sbottom doublet, which is the same in both realizations and reads (see e.g. Ref.~\cite{Fan:2014axa}) $\widehat{T}_{\mathrm{scalar}, A} \approx L_T  m_t^2 / (6 \widetilde{m}^2)$, where for convenience we have defined $L_T \equiv N_d y_t^2 / (16\pi^2)$. We now discuss, separately for the two models, the contributions of hidden sector loops.
\subsection{Tripled top model in this work}
The contribution of $B$ fermion loops reads $\widehat{T}_{\mathrm{fermions}, B} = \alpha N_d  \left\{ r_W - r_Z/ 2 \right\} / (16 \pi s_w^2 c_w^2)\,$, where $r_W - r_Z/2$ is the quantity in curly brackets in Eq.~(24) of Ref.~\cite{Anastasiou:2009rv}. Plugging in the explicit expressions of the couplings and masses, and expanding in the relevant limit $\omega \ll m_t \ll M$, we find $\widehat{T}_{\mathrm{fermions}, B+C} \approx  4 L_T m_t^2 / (3M^{2})$ where we have included an overall factor $2$ that sums over the $B$ and $C$ sectors. Turning to scalar loops, the ``$S^c$'' sector in Eqs.~\eqref{eq:Bscalarmasses_newmodel} and \eqref{eq:Bscalarmasses_newmodel_2} is supersymmetric, so its contribution has the same parametric scaling as $\widehat{T}_{\mathrm{fermions}, B+C}$ but a smaller numerical coefficient, and we neglect it. On the contrary, the contribution of the light scalars in the ``$S$'' sector is potentially important. We calculate it by exploiting the well-known fact that $\widehat{T} = (\delta Z_+ - \delta Z_3)_{\rm Landau\; gauge}\,$, where $\delta Z_{+, 3}$ are the wavefunction renormalizations of the charged and neutral Goldstones, respectively~\cite{Barbieri:2007gi,Orgogozo:2011kq}. The pieces of the $D$-term scalar potential that contain the Higgs field are
\begin{align} \label{eq:V_Dterms}
V_B \,=&\, y_t^2 ( | h^0 |^2 + | h^+ |^2) | \tilde{u}_B^c |^2 +  y_t M [ (h^{0\ast} \tilde{t}_B^{\prime c} + h^- \tilde{b}_B^{\prime c}) \tilde{u}_B^{c \ast} + \mathrm{h.c.} ]  \nonumber \\
\,+&\, y_t^2 |h^0 \tilde{t}_B - h^+ \tilde{b}_B |^2  + y_t \omega [ (h^0 \tilde{t}_B - h^+ \tilde{b}_B)^\ast \tilde{u}^\prime_B + \mathrm{h.c.}].
\end{align}
Focusing on the fields without the ``$c$'' superscript, rotating to the mass eigenbasis and taking $\omega \to 0$, we find that the relevant coupling is $- y_t m_t \widetilde{T}_B^\ast h^+ \tilde{b}_B + \mathrm{h.c.}$, which renormalizes $\delta Z_+$ and thus
\begin{align}
\widehat{T}_{S, B}  = L_T m_t^2 f [M^2_T, M^2_b] \quad \to \quad \widehat{T}_{S, B + C} \simeq \frac{1}{3} L_T \frac{m_t^2}{\Delta^2}\,,
\end{align}
where $f [m_A^2, m_B^2] \equiv ( m_A^4 - m_B^4 + 2m_A^2 m_B^2 \log m_B^2 / m_A^2 )/ [2 (m_A^2 - m_B^2)^3 ]$, while $M^2_T$ was defined in Eq.~\eqref{eq:scalar_masses} and $M^2_b = \Delta^2$. This is the dominant correction to $\widehat{T}$ in this model. Numerically, requiring $\widehat{T}_{S, B + C} \lesssim 10^{-3}$ gives $\Delta \gtrsim 400\;\mathrm{GeV}$, which can be seen as a rough current lower bound from EWPT on the masses of the EW-doublet scalar top partners. 

\subsection{Original tripled top model with singlet top partners}
The contribution of the hidden fermions, expanding for $m_t \ll \omega \ll M$, is $\widehat{T}_{\mathrm{fermions}, B+C} \approx L_T m_t^2 / M^{2}$ at the leading order. For the scalars, the roles of the $S$ and $S^c$ sectors are reversed with respect to the model considered in this paper. We neglect the ``$S$'' sector in Eqs.~(36) and (37) of Ref.~\cite{Cheng:2018gvu}, which is supersymmetric and whose contribution is therefore subleading to $\widehat{T}_{\mathrm{fermions}, B+C}$. On the other hand, the light scalars in the ``$S^c$'' sector have sizable mixing and the associated correction is a priori relevant. To calculate it, the starting point is Eq.~\eqref{eq:V_Dterms} with $M \leftrightarrow \omega$. Focusing on the fields with the ``$c$'' superscript, we rotate to the mass eigenbasis, expand $h^0 = \langle h^0 \rangle  + (h - i \pi_3) / \sqrt{2}$ and neglect quartic interactions, which at one loop give rise to tadpole diagrams that do not renormalize the Goldstone wavefunctions. We arrive then at the relevant couplings
\begin{equation}
y_t \omega \left[ \tfrac{ i }{\sqrt{2} } \pi_3 \tilde{s}_\Delta^{c \ast} \tilde{s}_\omega^c + h^+ \tilde{b}_B^{\prime c \ast} (\cos\phi_R \tilde{s}_\Delta^c + \sin\phi_R \tilde{s}_\omega^c )    \right] + \mathrm{h.c.} \,,
\end{equation}
from which we obtain
\begin{align} \label{eq:Tscalars_originalTT}
\widehat{T}_{S^c, B}  = L_T \,\omega^2 \left\{ \cos^2 \phi_R f [m^2_{\tilde{s}_\Delta^c}, m^2_{\tilde{b}_B^{\prime c}}] + \sin^2 \phi_R f [m^2_{\tilde{s}_\omega^c} , m^2_{\tilde{b}_B^{\prime c}} ] - f [m^2_{\tilde{s}_\Delta^c} , m^2_{\tilde{s}_\omega^c}] \right\}.
\end{align}
Expanding to leading order in $m_t^2$, we find
\begin{equation} \label{eq:T_originalTT}
\widehat{T}_{S^c, B + C} \approx \frac{1}{3}L_T m_t^2 \omega^2 \left[ \frac{ \omega^6 + 9\, \omega^4 \Delta^2 - 9\, \omega^2 \Delta^4 - 6 \, \omega^2 \Delta^2 (\omega^2 + \Delta^2) \log \omega^2 / \Delta^2 - \Delta^6}{(\omega^2 - \Delta^2)^5} \right] .
\end{equation}
Numerically, this correction is well below the current bound $\widehat{T}\lesssim 10^{-3}\,$: e.g., for $\omega = 500\;\mathrm{GeV}$ and $\Delta = 300$~GeV we have $\widehat{T}_{S^c, B + C} \simeq + 1.7 \times 10^{-4}$, computed using the full numerical expression in Eq.~\eqref{eq:Tscalars_originalTT}.

\section{Analysis of $Z\to b\bar{b}\nu \nu+X$ at $Z$ Factories}\label{App:bbnunu_Zfact}
\begin{figure}[t]
\includegraphics[width=0.48\textwidth]{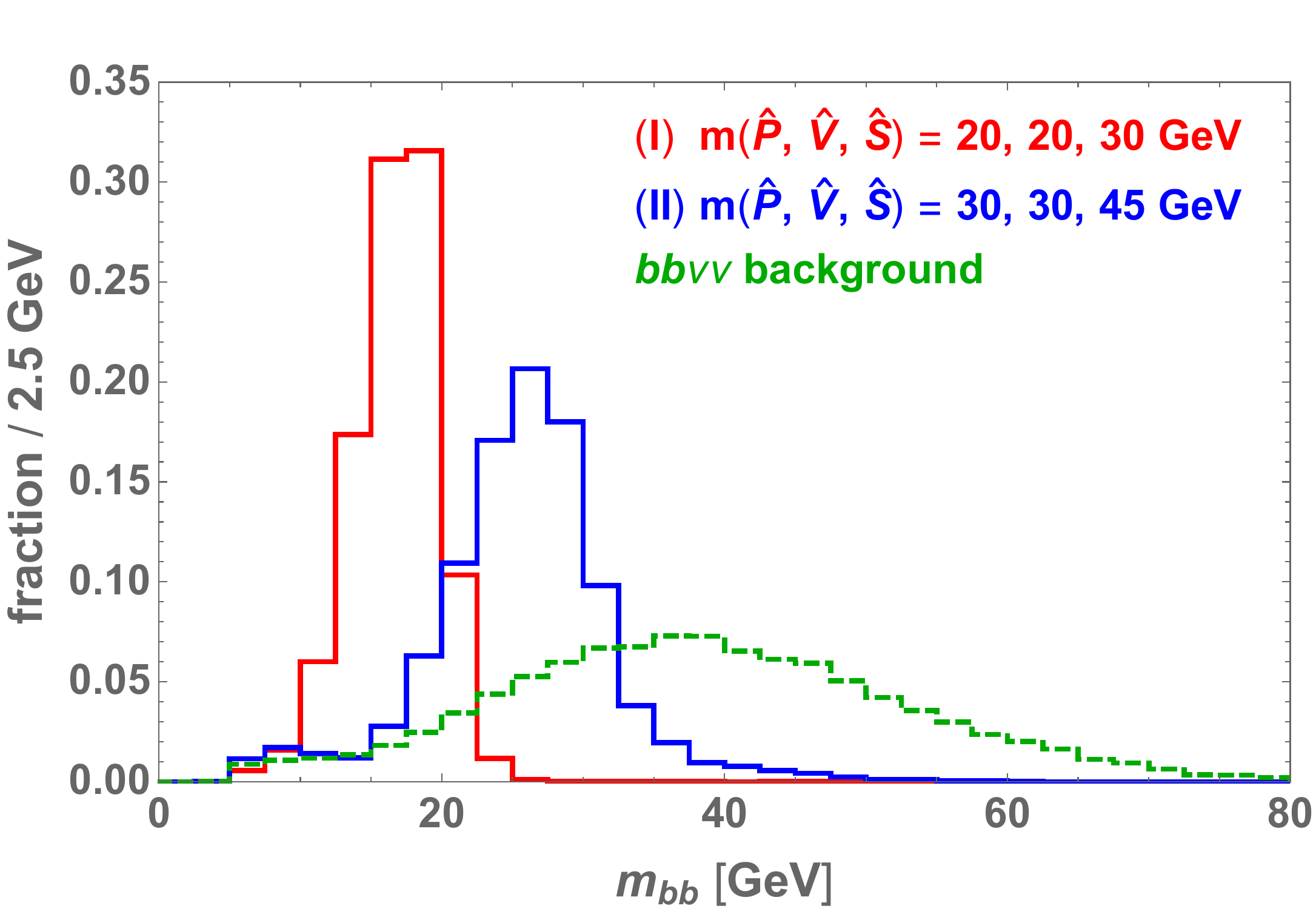}\hspace{2mm}
\includegraphics[width=0.48\textwidth]{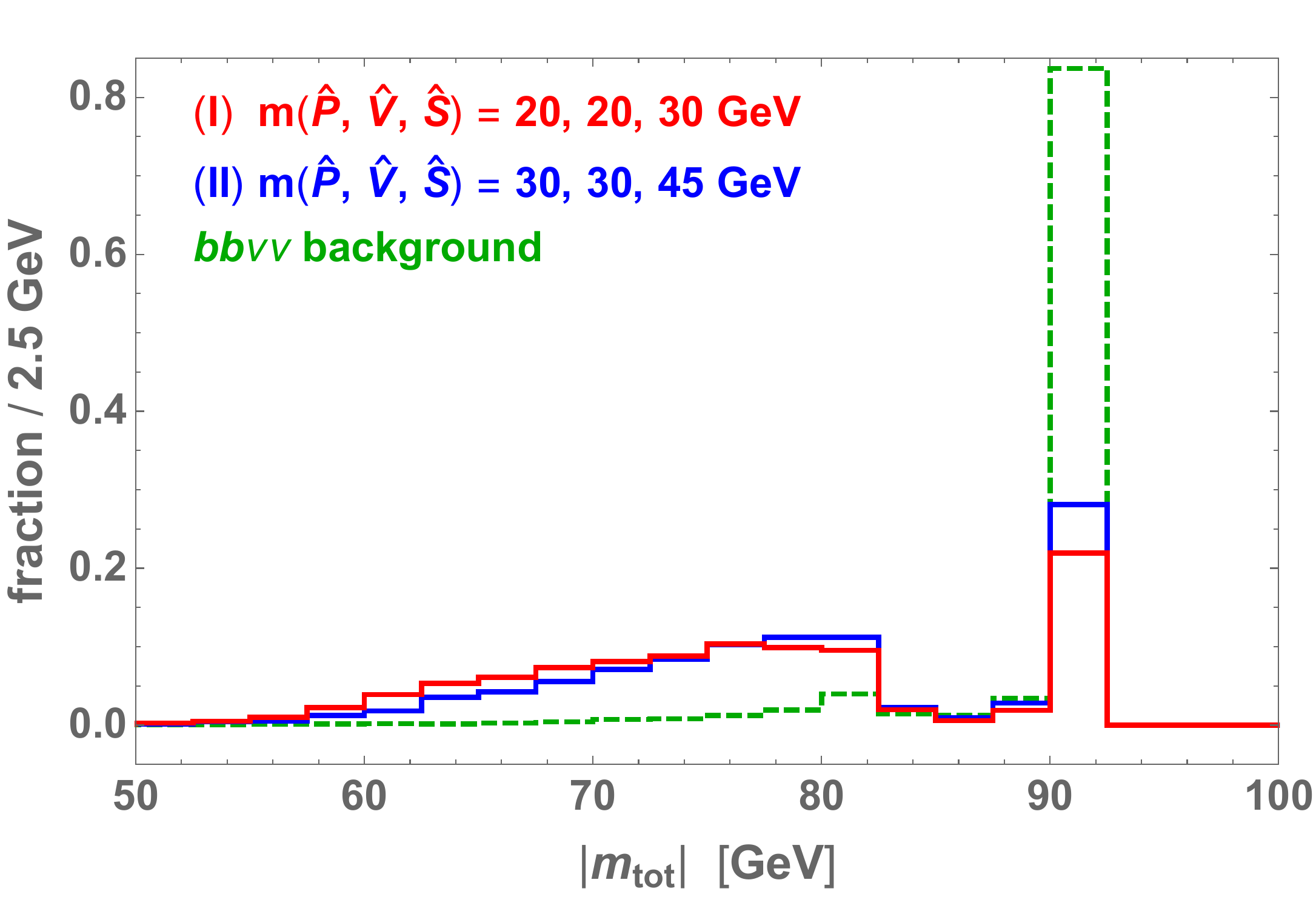}
\caption{\label{f.bbnunu_Zfact}Normalized distributions of the $b\bar{b}$ invariant mass ({\it left}) and absolute value of $m_{\rm tot}$ ({\it right}) for the $Z\to \omp \opp \to b\bar{b} \nu \nu + X$ signal and the irreducible $b\bar{b}\nu \nu$ background at a $Z$ factory, after the basic selection that requires two $b$-tagged jets.} 
\end{figure}

\begin{figure}[t]
\includegraphics[width=0.48\textwidth]{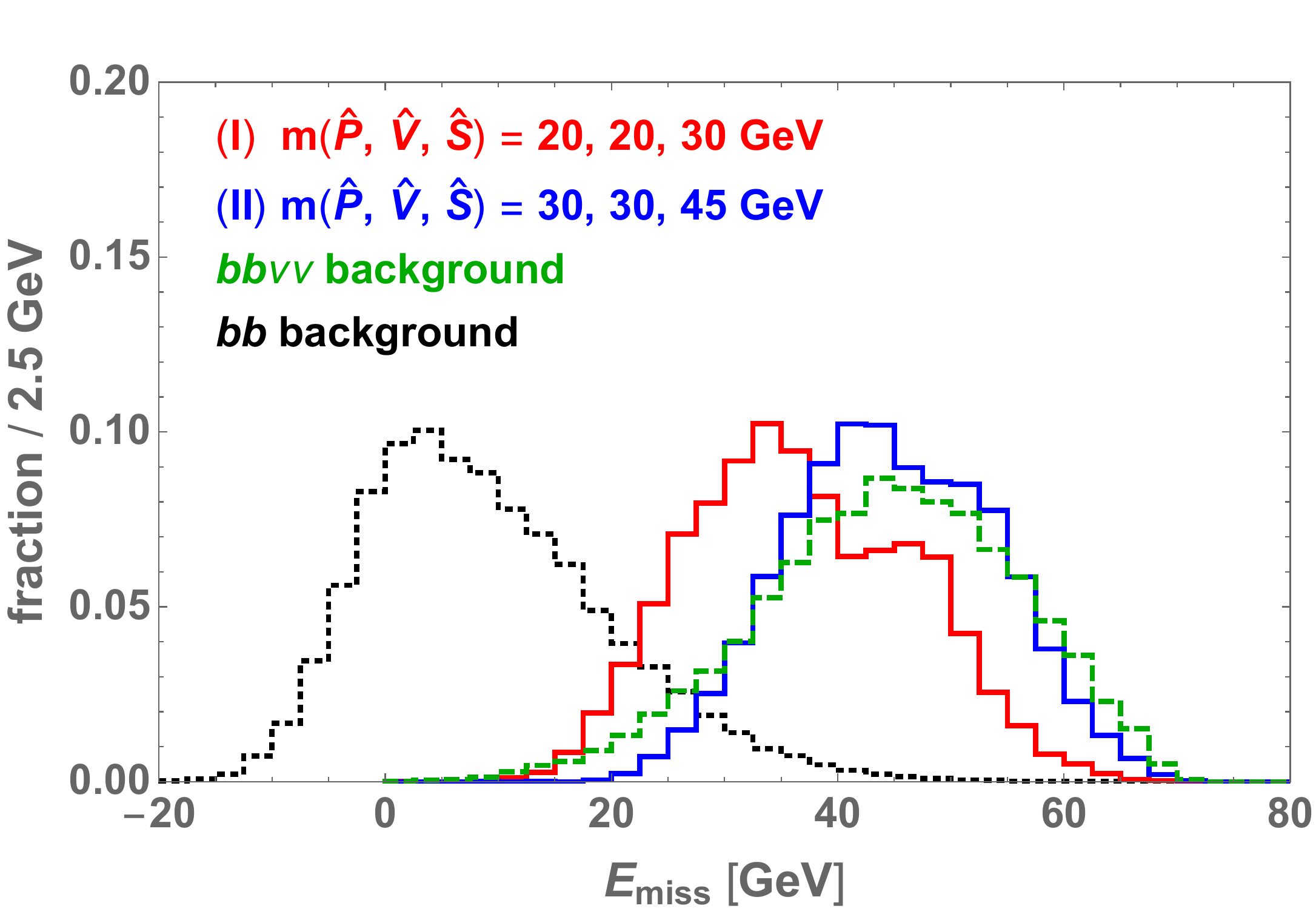}\hspace{2mm}
\includegraphics[width=0.48\textwidth]{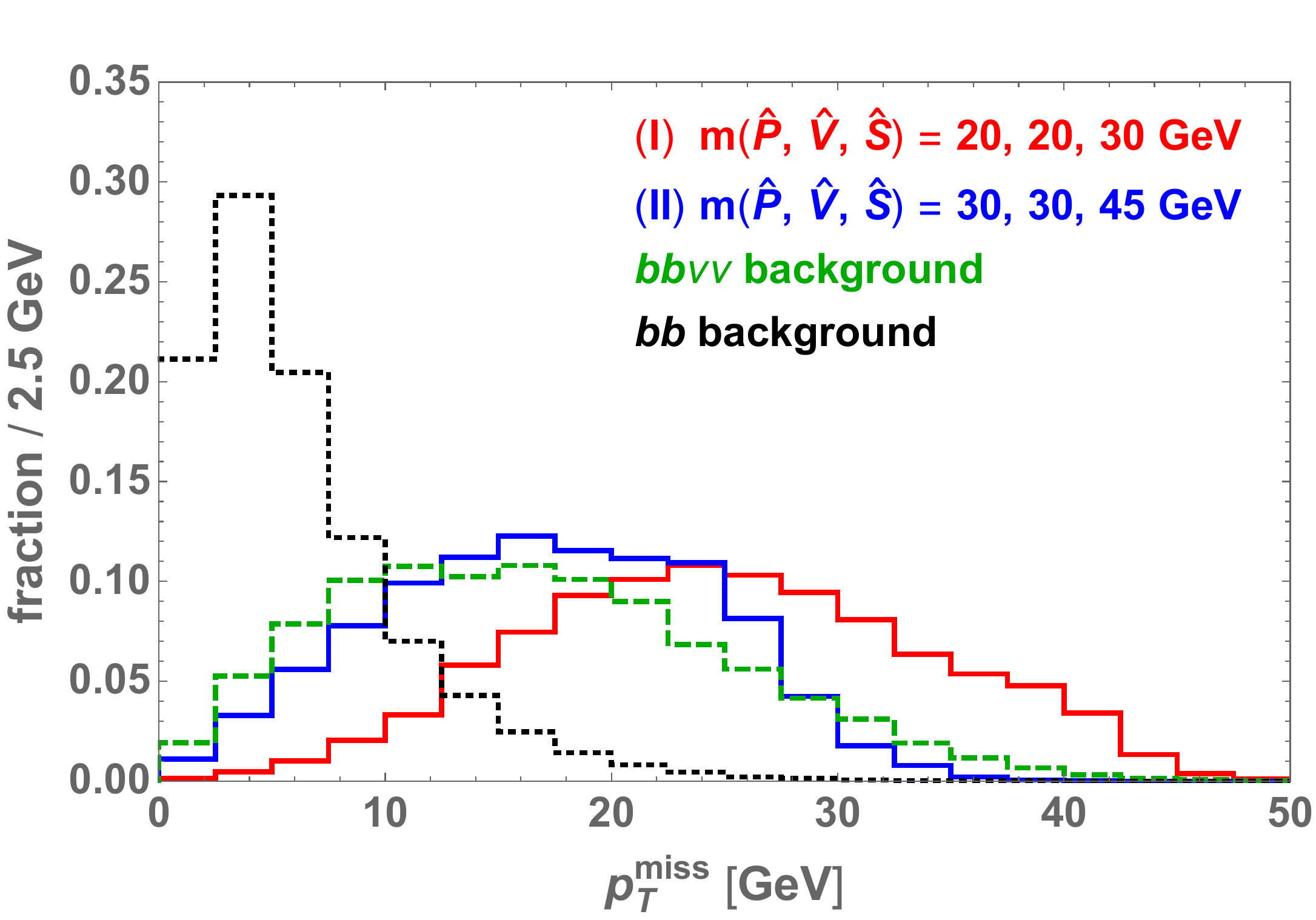}
\caption{\label{f.bbnunu_Zfact_redbkg}Normalized distributions of the missing energy ({\it left}) and missing transverse momentum ({\it right}) for the $Z\to \omp \opp \to b\bar{b} \nu \nu + X$ signal, the irreducible $b\bar{b}\nu \nu$ background and the reducible $b\bar{b}$ background at a $Z$ factory, after the basic selection requiring $2$ $b$-jets.} 
\end{figure}

The prompt signal $ee\to Z \to ( \omp \to b\bar{b} )(\opp  \to \lmm f\bar{f})$ at a $Z$-factory was discussed in Subsec.~\ref{sec:prompt}, focusing on $\lmm \to \ell \ell$. However, given the very clean collider environment it is also interesting to consider $\lmm \to \nu \bar{\nu}$, which benefits from the larger $\mathrm{BR}(\lmm \to \nu \bar{\nu}) \approx 0.20$. The resulting final state is $b\bar{b}\;+\,$missing momentum$\,+\,X$, for which the irreducible SM background is $ee \to b\bar{b}\nu\nu$ with amplitude at $O(g_w^4)$. As basic selection we simply require $2\, b\,\mbox{-}\mathrm{jets} \;\mathrm{with}\; E_b > 10\;\mathrm{GeV}$ and $|\eta_b| < 2.3$\,. Based on the normalized distributions for signal and background after the basic selection, shown in Fig.~\ref{f.bbnunu_Zfact}, we implement the cuts
\begin{equation}
m_{bb} \in [m_{\omp} - 10\;\mathrm{GeV}, m_{\omp} + 5\;\mathrm{GeV}]\,, \qquad | m_{\rm tot} | < 85\;\mathrm{GeV}.
\end{equation}
We have defined $m_{\rm tot} = \sqrt{(p_b^1 + p_b^2 + p_{\rm miss})^2}\,$, where the missing four-momentum is $p_{\rm miss}^\mu = (E_{\rm miss}, \vec{p}_{\rm miss})$ with $E_{\rm miss} = \sqrt{s} - \sum_i E_i$ ($\sqrt{s} = m_Z$) and $\vec{p}_{\rm miss} = - \sum_i \vec{p}_i\,$. The sums in the definitions of $E_{\rm miss}$ and $\vec{p}_{\rm miss}$ run over all reconstructed objects, in particular only jets with $p_T > 5\;\mathrm{GeV}$ are included.\footnote{The resulting $p_T^{\rm miss} = \sqrt{p_{\mathrm{miss}, x}^{\,2} + p_{\mathrm{miss}, y}^{\,2}}$ displays a slightly harder distribution compared to the $\slashed{E}_T$ provided by Delphes. The latter is not used in this analysis.} By design, if the only reconstructed objects in the event are the two $b$'s then $m_{\rm tot} = m_Z$ , so $m_Z - m_{\rm tot}$ is effectively a measure of additional activity in the event. 
Additionally we impose $E_{\rm miss} > 30\;\mathrm{GeV}$, $p_T^{\rm miss} > 20\,(10)\;\mathrm{GeV}$ for I~(II), and veto extra $b$-jets with $p_T^b > 5 \;\mathrm{GeV}$ and $| \eta_b | < 2.3$. These further requirements should suppress the reducible $b\bar{b}$ background, see Fig.~\ref{f.bbnunu_Zfact_redbkg}. The total acceptance times efficiency for the signal is $(\mathcal{A}\,\epsilon)_{\rm tot}^{\rm I, \,II} = 9.3\%, 8.8\%$, whereas the expected $bb\nu \nu$ background yield at TeraZ is of $130$ and $390$ events, respectively. The resulting $95\%$ CL bounds are, in terms of the branching ratio $\mathrm{BR}(Z\to \omp \opp \to b \bar{b} + \mathrm{invisible} + X)$, $3.3 \times 10^{-8} \, (2.2\times 10^{-10})$ at GigaZ~(TeraZ) for benchmark I, and $3.8 \times 10^{-8}\, (3.9\times 10^{-10})$ for benchmark II. Comparing to the results obtained in Ref.~\cite{Liu:2017zdh} for $Z\to \phi_d A^\prime \to ( b \bar{b} )( \chi \bar{\chi} )$ with $\chi$ an invisible particle, we find that our BR bounds are only slightly weaker (by a factor $\sim 2$ at TeraZ), even though our topology is somewhat different due to presence of an extra $f\bar{f}$ pair. Our GigaZ bounds on $M$ are
\begin{equation} \label{eq:Zfactory_nunull_bounds}
(\mathrm{I}) \quad M \gtrsim 6.6\;\mathrm{TeV} \left( \frac{ f_{\omp \opp} }{1}\right)^{1/4} , \qquad\quad (\mathrm{II}) \quad M \gtrsim 6.4\;\mathrm{TeV} \left( \frac{ f_{\omp \opp} }{1}\right)^{1/4}. \qquad (\mathrm{GigaZ})
\end{equation}
For the larger values of $M$ that can be probed at TeraZ, the $\omp$ becomes long-lived, which will require a different experimental approach. 

An important caveat is that the above constraints were obtained neglecting the reducible $b\bar{b}$ background, to illustrate the best sensitivity that can be achieved. With our simulation settings, however, we find that the selection we described in this Appendix can only partially suppress $b\bar{b}$, whose inclusive cross section is approximately $6.5\;\mathrm{nb}$. After all cuts, at TeraZ we estimate $\sim 3 \times 10^6$ remaining events for analysis I, and $\sim 5 \times 10^7$ events for analysis II. These yields are $10^{4\mbox{-}5}$ times larger than the irreducible ones, and if taken at face value they would lead to a severe degradation of our results. However, we expect that specifically designed cuts, a more refined modeling of the detector, and advanced analysis techniques will enable a further suppression of the $b\bar{b}$ contamination while preserving a high signal efficiency. A detailed analysis of this aspect is left for future work.

\section{Emerging Jet Search at $Z$ Factories}\label{App:EMJ_Zfactories}
In this Appendix we derive projected limits on BR($Z\to \overline{\psi}_{B,C}\psi_{B,C}$) from the search for emerging jets (EMJs) at $Z$ factories. We follow closely the recent CMS analysis of EMJs at the LHC~\cite{Sirunyan:2018njd}. For each jet, the variables $\left\langle IP_{2D} \right\rangle$ and $\alpha_{3D}$ are defined, calculated using the information of charged tracks with $p_T>1$~GeV. $\left\langle IP_{2D} \right\rangle$ is defined to be the median transverse impact parameter ($D_{xy}$) of the tracks, while $\alpha_{3D}$ is the sum of the $p_T$ of the tracks that belong to the primary vertex (PV), divided by the scalar $p_T$ sum of all tracks in the jet. Only tracks satisfying
\begin{equation}
\sqrt{\bigg(\frac{D_z}{0.01~{\rm cm}}\bigg)^2 + \bigg(\frac{D_{xy}}{\sigma(d)}\bigg)^2} < 4 \,,
\end{equation}
are assumed to originate from the PV, where $\sigma(d)=\sqrt{0.003^2+(0.001\, p_{T(\rm track)}/\mathrm{GeV})^2}$~cm is the uncertainty of $D_{xy}$~\cite{Chatrchyan:2014fea}. In order to validate our approach, we first try to reproduce the EMJ-1 benchmark results in Ref.~\cite{Sirunyan:2018njd}. The LHC tracker resolutions of $D_{xy}$ and $D_{z}$ are taken from Refs.~\cite{ATLAS:2012jma,Chatrchyan:2014fea}. The EMJ signal and ($b$-)jet pair backgrounds are simulated using $Z^\prime$ decays with $M_{Z^\prime}=250$~GeV, and each jet is required to have $p_T>50$~GeV and $|\eta|<2$. Applying the \mbox{EMJ-1} selection we obtain a light jet misidentification rate $\epsilon_j \simeq 2\times10^{-3}$ for track multiplicity $\in [6,10]$ and $4\times 10^{-4}$ for track multiplicity $\in [11,16]$, which are close to those reported in Ref.~\cite{Sirunyan:2018njd}.  Similarly, for $b$-jets we find $\epsilon_b \simeq 4\times10^{-2}$ and $2\times 10^{-3}$ for the lower- and higher-track multiplicity bins. However, at the LHC a search for only $Z\to~$2 EMJs is not feasible due to trigger requirements, and we must resort to production in association with additional particles, such as $ZZ \to 2$ EMJs$\,+\,2\ell$ or \mbox{$Zj \to 2$ EMJs$\,+\,$hard jet}. The resulting sensitivity on the $Z$ branching ratio to the hidden sectors is at the level of $10^{-(\operatorname{4-5})}$, weaker than the LHCb reach described in Sec~\ref{sec:emerging}.

At a future $Z$ factory, the search reach will depend on the detector resolution. For a conservative estimate, in our analysis we use the LHC tracker resolution described above. The EMJ signal and $Z$-pole di($b$-)jet backgrounds are simulated using Pythia8, with the same benchmark parameters adopted in the LHCb analysis of Sec.~\ref{sec:emerging} (in particular, $m_{\omp} = m_{\lmm} =1$~GeV). We require $p_T>5$~GeV and $|\eta|<2$ for each jet. Since $\lmm$ has a $\sim 40\%$ branching ratio to neutrinos and the $\omp$ lifetime is very long, a significant fraction of the EMJ energy is carried by invisible particles, reducing the signal efficiency. On the other hand, the $\lmm$ has a $\sim 6\%$ branching ratio to $ee$ and $\mu\mu$ pairs, which are rare in SM jets. The lepton tracks can thus be used to better separate the EMJs from the SM jets. 
\begin{figure}[t]
\includegraphics[scale=0.43]{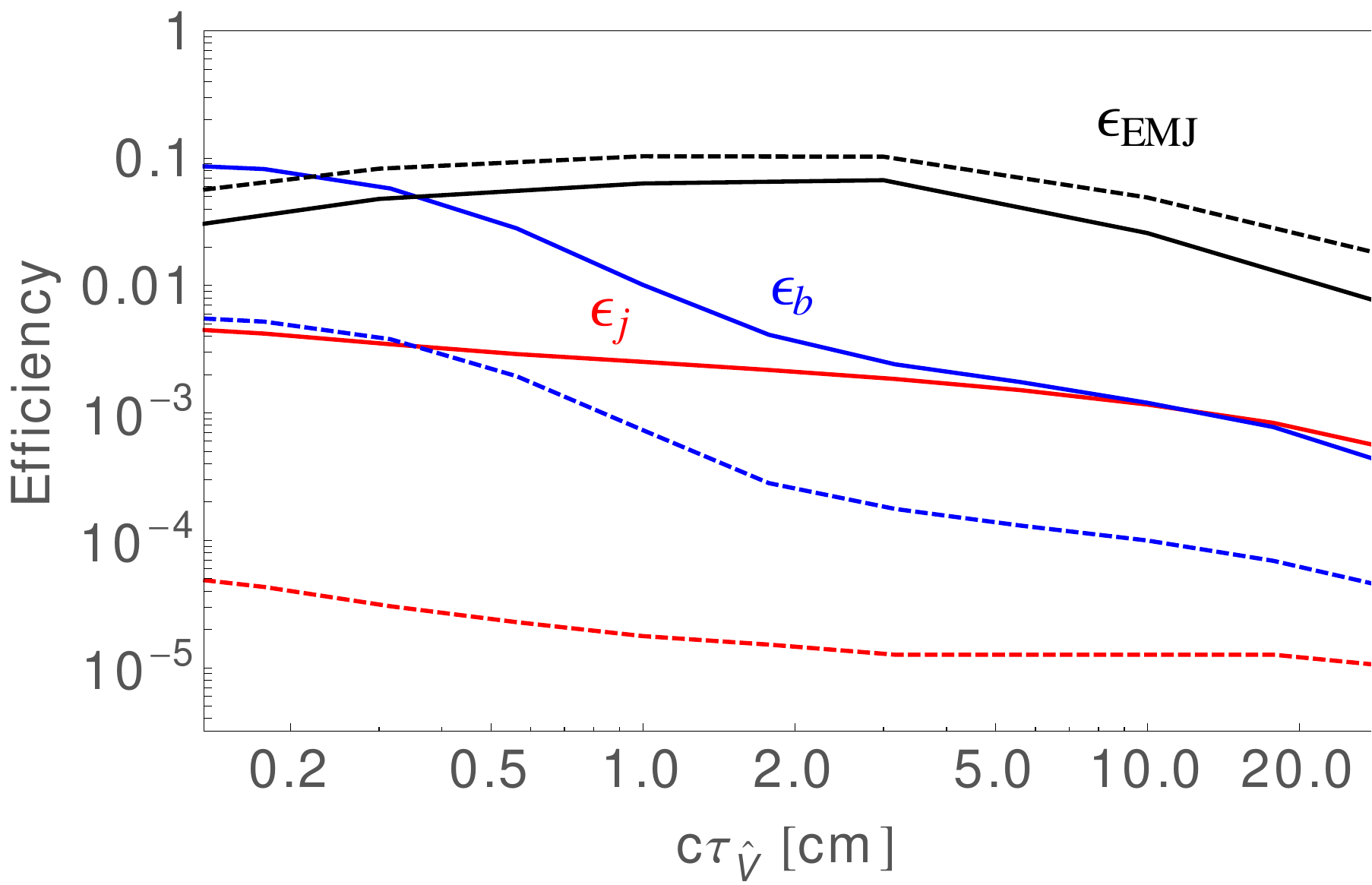}\hspace{1mm}
\includegraphics[scale=0.43]{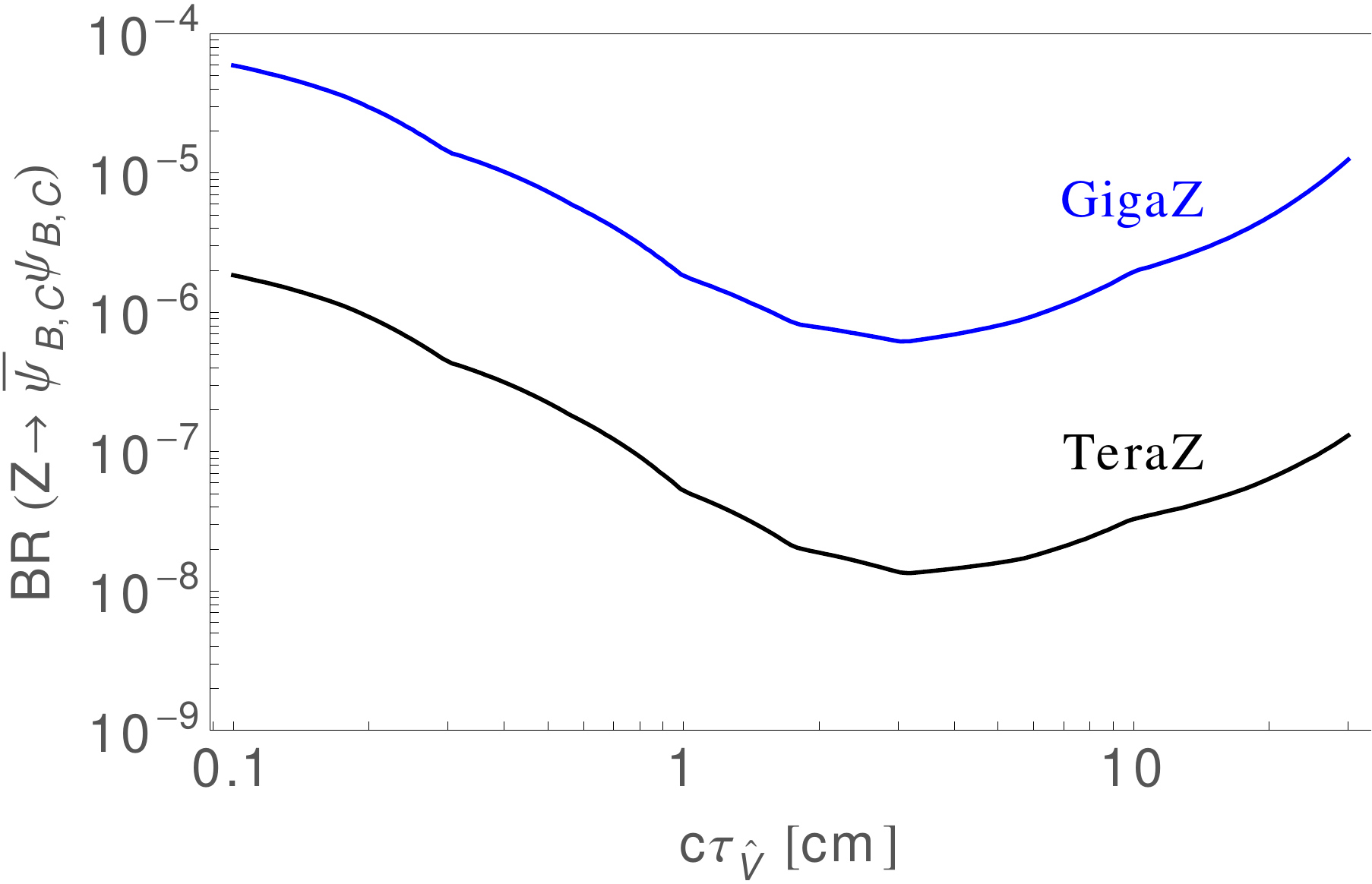}
\caption{{\it Left: } Emerging jet signal efficiency (black), $b$-jet misidentification rate (blue) and light-jet misidentification rate (red) as function of the target $c\tau_{\lmm}$, for the $Z$-pole emerging jet search. Solid curves correspond to the requirements $N_{\rm track}\geq 5$ and 0 or 1 lepton tracks. Dashed curves correspond to $N_{\rm track}\geq 3$ and $\geq 2$ lepton tracks. {\it Right: } Projected $95\%$ CL constraints on the $Z$ branching ratio to the hidden sectors.}
\label{fig:emergingjet_Z}
\end{figure}
For a target $c\tau_{\lmm}$, each jet is required to satisfy $\left\langle IP_{2D} \right\rangle \in [c\tau_{\lmm}/10,25\;\mathrm{cm}]$ and $\alpha_{3D}<0.25$. To avoid backgrounds from long-lived SM hadrons, such as $K_S$~\cite{Schwaller:2015gea}, only EMJ candidates with track multiplicity $N_{\rm track}\geq 5$ are accepted if they contain 0 or 1 lepton track. The cut on $N_{\rm track}$ is loosened to $\geq3$ if an EMJ candidate has $\geq 2$ lepton tracks. We plot the emerging jet efficiency and SM jet misidentification rates versus the target $c\tau_{\lmm}$ in the left panel of Fig.~\ref{fig:emergingjet_Z}, where the solid~(dashed) curves correspond to the case with $0$ or $1$~($\geq 2$) lepton track(s). Lepton tracks from $\lmm \to \ell\ell$ decays clearly help to discriminate the EMJs from the SM backgrounds, especially in the case of jets originating from light SM quarks. Due to its higher efficiency and lower misidentification rates, the lepton-track-rich channel provides the leading sensitivity; the largest background comes from SM $Z\to b\bar{b}$ decays. The combined limits on the $Z$ branching ratio are shown in the right panel of Fig.~\ref{fig:emergingjet_Z}. For $c\tau_{\lmm} \sim \mathrm{few}\; \mathrm{cm}$ the sensitivity reaches $\mathrm{BR}(Z\to \overline{\psi}_{B,C}\psi_{B,C})\sim 10^{-6}\,(10^{-8})$ at GigaZ~(TeraZ). Improved detectors and the use of information from calorimeters or the muon system may further extend the reach. Emerging jet searches at the LHC may also benefit from LLP-specific triggers and advanced analysis strategies. The study of these topics is beyond the scope of this paper, and is deferred to future work.

\vspace{4mm}
\bibliography{tripletPhenobib}

\end{document}